\documentstyle[12pt,fleqn]{article}

\def \virg{\;\;,}
\def \point{\;\;.}
\def \up{\uparrow}
\def \down{\downarrow}
\def \ve#1{\mbox{\boldmath $#1$}}

\def \fref#1{fig.\ref{#1}}

\def\cambaselines{\baselineskip=10.10pt
		  \lineskip=0pt
		  \lineskiplimit=0pt}
\def\oneskip{\vskip\baselineskip}
\cambaselines
\def\endmode{\par\endgroup}
\def\title{\begingroup \raggedright \noindent  \Large \bf }
\def\endtitle{\endmode \oneskip\oneskip\oneskip\oneskip}
\def\authors{\begingroup \parindent=2.0truecm \obeylines }
\def\endauthors{\endmode \oneskip\oneskip}

\def\epsffile#1{{}}
\def\rotate#1{{}}
\newlength{\epsfysize}

\newcommand{\sectio}[1]{\section{#1}\setcounter{equation}{0}}

\begin{document}
\oneskip \oneskip \oneskip

\title

THE METAL--INSULATOR TRANSITION IN ONE DIMENSION

\endtitle


\authors

H.J. SCHULZ
Laboratoire de Physique des Solides
Universit\'e Paris--Sud, 91405 Orsay, France

\endauthors

\noindent {\bf Abstract:} The low--energy excited states of a system of
interacting one--dimensional fermions in a conducting state are collective
charge and spin density oscillations. The unusual physical properties of
such a system (called ``Luttinger liquid'') are characterized by the
velocities $u_\rho$ and $u_\sigma$ of the charge and spin excitations, as
well as by a parameter $K_\rho$ that determines the power law behavior of
correlation functions. Umklapp scattering occuring at half--filling or other
commensurate bandfilling can lead to a transition into an insulating state,
characterized in particular by a gap in the charge excitations (the
Mott--Hubbard gap). The properties in the vicinity of the transition are
shown to depend on both the way the transition is approached (constant
bandfilling and varying interaction, or constant interaction and varying
bandfilling) and on the ``order'' of the commensurability. In particular,
even and odd fractional fillings show quite different behavior. This
behavior is illustrated in detail using lattice models like the Hubbard
model and its extensions.

\sectio{Introduction}
A theoretical understanding of interacting fermion systems in one
dimension is important for a number of reasons. On the one hand, in the
physics of quasi-one-dimensional organic conductors
\cite{jerome_revue_1d} or of conducting polymers
\cite{heeger_revue_polymeres} interaction
effects play a major role. On the other hand, one--dimensional models
can be easier to understand than their higher-dimensional versions, or
even exactly solvable,
as is the case with the prototypical model of correlated fermions,
the Hubbard model \cite{lieb_hubbard_exact}. They therefore can provide
valuable information on the role of correlation effects in higher
dimension, e.g. on the physics of correlated fermions in two dimensions
which is thought to be at the origin of the many interesting properties
of high-temperature superconductors
\cite{anderson_hgtc_hubbard,anderson_hgtc_cours}.

Theoretical work on interacting fermions in one dimension has progressed
along a number of different lines. One approach has been the perturbative
investigation of the weak coupling limit. Even this is in fact not entirely
straightforward, mainly because of the infrared divergences encountered in
this type of calculation which require a renormalization group treatment. A
complete review of this approach has been given by S\'olyom
\cite{solyom_revue_1d}. An alternative and more general approach is provided
by the so--called ``bosonization'' method, which is based on the
equivalence between interacting fermions and
noninteracting bosons (representing density fluctuations) and on the
expression of fermionic operators in terms of these bosons. Combined with
the renormalization group approach, the bosonization method provides a
rather straightforward description of the peculiar properties
of one--dimensional
interacting fermion systems (``Luttinger liquid''), and one finds
 that the low--energy physical
properties are determined by only three parameters: the velocities of
collective charge-- and spin--density oscillations ($u_{\rho,\sigma})$,
and a coefficient $K_\rho$ that determines the long--distance decay of
correlation functions. These coefficients play a role
similar to the Landau
parameters of (three--dimensional) Fermi liquid theory. A number of
physical properties depending on these parameters are discussed below, but
let us mention here that in particular the coefficient $K_\rho$ is
important in a much wider variety of phenomena:
the temperature dependence of the NMR
relaxation rate \cite{bourbonnais_rmn_2} or of X-ray scattering
intensities \cite{pouget_exponents}, the energy dependence of
photoemission spectra \cite{dardel_photo}, the effect
of impurities \cite{giamarchi_loc}, or possible low-temperature ordered
states in systems of coupled chains \cite{bourbon_couplage} all depend on
it. A brief discussion
of bosonization will be given in chapter \ref{wcsec}, but for  more
detailed and rigorous derivations and results, the reader is referred to
more specialized articles \cite{emery_revue_1d,haldane_bosonisation}.

As presented in sec. \ref{wcsec}, the theory applies to translational
invariant systems, and therefore one always has a conducting state. On the
other hand, if one considers lattice systems, umklapp scattering breaks
translational invariance. In sec. \ref{umsec}, I will show how this type of
scattering can be incorporated into the bosonization formalism. One then
finds insulating phases of the Mott--Hubbard type, with properties
depending in an interesting way on the order of commensurability.
There are two types of metal--insulator transitions: either at constant
particle density, as a function of interaction strength, or at constant
interaction, as a function of particle density. The ``critical behavior'' of
the two types of transitions is quite different.

A rather different approach (at least until recently) is based on the famous
``Bethe ansatz'' \cite{bethe_xxx} which in particular has made possible an
exact solution of both continuum fermions interacting via $\delta$--function
potentials \cite{gaudin_fermions,yang_fermions} and of the one--dimensional
Hubbard model \cite{lieb_hubbard_exact} (and of many other interesting
models). In section \ref{hubsec} I will discuss the low--lying excitations
of the one--dimensional Hubbard model as obtained from the exact solution.
This will give a rather concrete illustration of the
concept of ``holons'' and ``spinons''. Moreover, one observes interesting
changes as the metal--insulator transition is approached.

The exact Bethe ansatz eigenfunctions are so
complicated that the direct calculation of correlation functions
and many other physical properties of the
one--dimensional Hubbard model is difficult
even for very small systems \cite{ogata_inf} and impossible in the
thermodynamic limit. In section \ref{corrsec} I present a method
\cite{schulz_hubbard_exact}
that allows in particular a determination of the coefficient
$K_\rho$ for arbitrary correlation strength. One then obtains a
rather detailed {\em and exact} description of the
low--energy (and low temperature) properties
and also of the metal--insulator transition occurring when the
average particle number per site, $n$, approaches unity. The method
generalizes rather straightforwardly to other models, where however
in general one has to rely on exact calculations for small systems and an
extrapolation to the thermodynamic limit.

\sectio{Luttinger liquids}
\label{wcsec}
\subsection{The spinless case}
\subsubsection{The model and its solution}
Let us consider the standard case of an interacting fermion model with
the Hamiltonian consisting of two parts: the kinetic energy and the
interaction. Omitting spin for the moment,
 the kinetic energy term is of the form
\begin{equation}
\label{h00}
H_0 = \sum_{k} \varepsilon_k c_{k}^\dagger c_{k} \virg
\end{equation}
where $c_{k}$ and $c_{k}^\dagger$ are the standard annihilation and
creation operators for a fermion with momentum $k$, and
$\varepsilon_k$ is the single--particle bandstructure. In a simple
tight--binding model one would have $\varepsilon_k = -2t \cos k$ (the
lattice constant is set to unity), but the precise form of $\varepsilon_k$
is unimportant here. The Fermi surface
consists just of the two points $\pm k_F$.

For weak interactions between the
particles, only states in the immediate vicinity of the Fermi points are
important. For these states, one then can linearize the electronic
dispersion
relation around the Fermi points, and the kinetic energy term takes the form
\begin{equation}
\label{h0}
H_0 = v_F \sum_{k} \{(k-k_F) a_{k}^\dagger a_{k} +
 (-k-k_F) b_{k}^\dagger b_{k} \}
\point
\end{equation}
Here the $a$ ($b$) operators refer to states in the vicinity of $+k_F$
($-k_F$), i.e. the $a$--particles move to the right, the $b$--particles move
to the left.
The $k$--summation is limited to an interval $[-k_0,k_0]$ around $k_F$
(typically, $k_0 \approx \pi/2$, but the precise value isn't important
here). The Fermi velocity is given by
\begin{equation}
v_F = \left. \frac{\partial \varepsilon_k}{\partial k} \right|_{k_F} \virg
\end{equation}
and the density of states  is $N(E_F) = 1/(\pi v_F)$.
In the {\em Luttinger model}, one generalizes this kinetic energy by
letting the cutoff $k_0$ tend to infinity. There then are two branches
of particles, ``right movers'' and ``left movers'', both with
unconstrained momentum and energy. At least for weak interaction, this
addition of extra states far from the Fermi energy is not expected to
change the physics much. However, this modification makes the model
exactly solvable even in the presence of nontrivial interactions.
Moreover, and most importantly, many of the features of this model carry
over even to strongly interacting fermions on a lattice.

We now introduce interactions between the fermions. As long as only
forward scattering of the type
$(k_F;-k_F) \rightarrow
(k_F;-k_F)$ or $(k_F;k_F) \rightarrow  (k_F;k_F)$ is introduced, the
model remains exactly solvable. Introducing the Fourier components of
the particle density operator for right and left movers by
\begin{equation}
\rho_+(q) = \sum_k a^\dagger_{k+q} a_k \;\; , \;\;
\rho_-(q) = \sum_k b^\dagger_{k+q} b_k \;\; ,
\end{equation}
the interaction Hamiltonian describing these processes takes the form
\begin{equation}
\label{hint}
H_{int}  =  \frac{1}{2L} \sum_{q} \{2
g_2(q) \rho_+(q) \rho_-(-q) +
g_4(q) [\rho_+(q) \rho_+(-q) + \rho_-(-q) \rho_-(q)] \}
\point
\end{equation}
Here, $g_2(q)$ and $g_4(q)$ are the Fourier transforms of a real space
interaction potential, and in a realistic case one would of course have
$g_2(q) = g_4(q)$, but it is useful to allow for differences between
$g_2$ and $g_4$.
For Coulomb
interactions one expects $g_2, g_4 > 0$.
In principle, the long--range part of the Coulomb repulsion  leads to a
singular $q$--dependence. Such singularities in the $g_i$ can be handled
rather straightforwardly and can lead to interesting physical effects
\cite{schulz_wigner}, but here I shall limit myself mainly to nonsingular
$g_2, g_4$.  Electron--phonon
interactions can lead to effectively attractive interactions between
electrons, and therefore in the following I will not make any restrictive
assumptions about the sign of the constants. One should however notice that
a proper treatment of the phonon dynamics and of the resulting retardation
effects requires more care \cite{voit_phonon}.

The model defined by eqs. (\ref{h0}) and (\ref{hint}) can be solved
exactly. The solution is based on the following facts:
\begin{enumerate}
\item the density fluctuation operators $\rho_{\pm}$ obey Bose type
commutation relations:
\begin{equation}
[\rho_+(-q),\rho_+(q')] = [\rho_-(q),\rho_-(-q')]
= \delta_{qq'} \frac{qL}{2\pi} \; \; , \quad
[\rho_+(q),\rho_-(q')] = 0 \point
\end{equation}
Moreover, for $q>0$ both $\rho_+(-q)$ and $\rho_-(q)$ annihilate the
noninteracting groundstate. These properties are closely related to the
existence of an infinity of states at negative energies and would not
have been true in a lattice system like the one described by
(\ref{h00}).
\item The kinetic part of the Hamiltonian can be re--written as a term
bilinear in boson operators, i.e. quartic in fermion operators:
\begin{equation}
\label{h0b}
H_0 = \frac{2 \pi v_F}{L} \sum_{q>0}
[\rho_+(q) \rho_+(-q) + \rho_-(-q) \rho_-(q)]   \point
\end{equation}
This equivalence may be made plausible noting that $\rho_+(q)$ creates
particle--hole pairs that all have total momentum $q$. Their energy is
$\varepsilon_{k+q} - \varepsilon_k$, which, because of the linearity of
the spectrum equals $v_F q$, {\em independently of} $k$. Thus, states
created by  $\rho_+(q)$ are linear combinations of individual
electron--hole excitations all with the same energy, and therefore are
also eigenstates of (\ref{h0}).
\item The states created by repeated application of $\rho_{\pm}$ on the
ground state form a complete set of basis states
\cite{haldane_bosonisation,heidenreich_bosonisation}.
\end{enumerate}
Putting together (\ref{h0b}) and (\ref{hint}), the complete interacting
Hamiltonian then becomes a bilinear form in boson operators, that is
easily diagonalized by a Bogolyubov transformation. A first consequence
is the expression for the excitation spectrum
\begin{equation}
\label{omq}
\omega(q) = |q| [(v_F + g_4(q)/(2\pi))^2 - (g_2(q)/(2 \pi))^2]^{1/2}
\point
\end{equation}
The diagonal boson operators are linear combinations of the original
$\rho$ operators, and consequently, these elementary excitations are
collective density oscillations, their energy being determined both by
the kinetic energy term and the interactions.

We note here that in order for the Bogolyubov transformation to be a
well--defined unitary transformation, $g_2(q)$ has to decrease at large
$q$ at least as $|q|^{-1/2}$. on the other hand, the large--$q$
behavior of $g_2$ is unimportant for the low--energy properties of the
model. We therefore in the following will almost always use a
$q$--independent $g_2$ and $g_4$. An approximate and frequently used way to
cure the divergences arising due to this procedure is to keep the
parameter $\alpha$ in subsequent formulae as a finite short--distance
cutoff, of the order of a lattice spacing. One can then also include the
``backward scattering'' $(k_F;-k_F) \rightarrow (-k_F;k_F)$, because for
spinless electron this is just the exchange analogue of forward scattering
and does not constitute a new type of interaction.

Up to this point, this construction does not allow for a direct
calculation of correlation functions like the one--particle Green's
function or more generally any function involving individual creation or
destruction operators. This type of correlation function becomes
tractable by representing single particle operators in terms of the
boson operators; To this end, we introduce the field operators
\begin{eqnarray}  \label{phi0}
\phi(x) & = & -\frac{i\pi}{L}\sum_{p \ne 0} \frac1{p}
e^{-\alpha |p| /2 -ipx} [\rho_+(p) + \rho_-(p)] - N \frac{\pi x}{L}
\virg \\
\label{Pi0}
\Pi(x) & = & \frac{1}{L}\sum_{p \ne 0}
e^{-\alpha |p| /2 -ipx} [\rho_+(p) - \rho_-(p)] + J /L
\point
\end{eqnarray}
Here $N$ and $J$ are the number of particles added to the ground state
and the difference between the number of right and left--moving
particles, respectively, and $\alpha$ is a cutoff parameter which
(at least in principle, see the discussion above) has to
be set to zero in the end of any calculation. $\phi$ and $\Pi$ obey
canonical boson commutation, relations:
\begin{equation}
[\phi(x),\Pi(y)] = i\delta (x-y) \virg
\end{equation}
and $\phi$ is related to the local particle density via
$\partial \phi / \partial x = - \pi \rho(x)$. The expression for the single
fermion operators then is
\begin{equation} \label{singlepsi0}
\psi_\pm(x) = \lim_{\alpha \rightarrow 0} \frac1{\sqrt{2\pi
\alpha}} \exp \left[
\pm i k_F x \mp i \phi(x) - i \theta(x) \right]
\virg
\end{equation}
where $\theta(x) = \pi \int^x \Pi(x') dx'$, and
the upper and lower sign refer to right-- and left--moving electrons,
respectively.
A detailed derivation of this important relation as an operator identity
is given in the literature
\cite{haldane_bosonisation,heidenreich_bosonisation}.
However, a simple plausibility argument can be given:
 creating a particle at site $x$ means
introducing a kink of height $\pi$ in $\phi$, i.e. at points on the left
of $x$ $\phi$ has to be shifted by $\pi$. Displacement operators are
exponentials of momentum operators, and therefore a first guess would be
$\psi^\dagger(x) \approx \exp (i \pi \int_{-\infty}^x \Pi(x') dx')$.
 However, this operator commutes with
itself, instead of satisfying canonical anticommutation relations.
Anticommutation is achieved by multiplying with an operator, acting at
site $x$, that changes sign each time a particle passes through $x$.
Such an operator is $\exp (\pm i \phi(x))$. The product of these two factors
then produces (\ref{singlepsi}).

The full Hamiltonian can also be simply expressed in terms of $\phi$ and
$\Pi$. Neglecting the momentum dependence of the $g_i$, one easily finds
\begin{equation}
\label{hbos0}
H = H_0 + H_{int}
= \int dx \left[ \frac{\pi u K}{2} \Pi(x)^2
       + \frac{u}{2\pi K} (\partial_x \phi )^2 \right] \point
\end{equation}
This is obviously just the Hamiltonian of an elastic string, with the
eigenmodes corresponding to the collective density fluctuations of the
fermion liquid. It is important to notice that these collective modes are
the only (low--energy) excited states, and that in particular {\em there
are no well--defined single particle excitations.} The parameters in
(\ref{hbos0}) are given by
\begin{equation} \label{uk0}
u = [(v_F+g_4/(2\pi))^2 - g_2^2/(2\pi)^2]^{1/2} \virg \;
K = \left[ \frac{2\pi v_F +g_4 - g_2}{2\pi v_F +g_4 + g_2}\right]^{1/2}
\point
\end{equation}
The energies of the eigenstates are $\omega(q) = u |q|$, in agreement
with eq. (\ref{omq}).

\subsubsection{Physical properties}
The simple form of the Hamiltonian (\ref{hbos0})
 makes the calculation of physical properties rather
straightforward. First note that acoustic phonons in one dimension have
a
linear specific heat. Consequently, the low-temperature specific heat of
interacting fermions is $C(T) = \gamma T$, with
\begin{equation}       \label{gamma0}
\gamma/\gamma_0 =  v_F/u  \point
\end{equation}
Here $\gamma_0$ is the specific heat coefficient of noninteracting
electrons of Fermi velocity $v_F$. Also,
 the coefficient $u / K$
fixes the energy needed to change the particle density, and consequently the
renormalization of the compressibility $\kappa$ is given by
\begin{equation}  \label{kappa0}
  \kappa / \kappa_0 = v_F K / u \virg
\end{equation}
where  $\kappa_0$ is the
compressibility of the noninteracting case.

The quantity $\Pi$ is proportional to the current density.
Obviously, the Hamiltonian commutes with the total current, and
therefore the frequency dependent conductivity is a delta function at
$\omega = 0$. Using the Kubo formula, one straightforwardly finds
\begin{equation}  \label{sig00}
\sigma (\omega) =  K u \delta (\omega ) \virg
\end{equation}
i.e. the product $K u $ determines the weight of the dc peak
in the conductivity.

The above properties
 are those of an ordinary Fermi liquid, the
coefficients $u$, and $K$ determining
renormalizations with respect to noninteracting quantities. We will now
consider quantities which show that a one--dimensional interacting
fermion system is {\em not a Fermi liquid}. Consider the
single--particle Green's function which can be calculated using the
representation (\ref{singlepsi0}) of fermion operators:
\begin{eqnarray}
\nonumber
G^R(x,t) & = & -i \theta(t) \langle
[\psi_{+}(x,t),\psi_{+}^\dagger(0,0)]_+ \rangle \\
\label{g0}
& = & -\frac{\theta (t)}{\pi} e^{ik_Fx} {\rm Re}
\left\{ \frac{1}{u t -x}
  \left[\frac{\alpha^2}{(\alpha+iu t)^2 + x^2} \right]^{\delta/2}
   \right\}
\point
\end{eqnarray}
 One then finds
for the momentum distribution function in the vicinity of $k_F$:
\begin{equation}
n_k \approx n_{k_F} -  {\rm const.}
 {\rm sign}(k-k_F) |k-k_F|^\delta \;\; \virg
\end{equation}
and for the single-particle density of states: $N(\omega ) \approx
|\omega |^\delta$, with $\delta=(K + 1/K -2)/4$, and $\beta$
is a model--dependent constant. Note that for any
$K \neq 1$, i.e. {\em for any nonvanishing interaction}, the
momentum distribution function and the density of
states have power--law singularities at the Fermi level, with a
vanishing single particle density of states at $E_F$. This
behavior is obviously quite different from a standard Fermi liquid
which would have a finite density of states and a step--like singularity
in $n_k$. The absence of a step at $k_F$ in the momentum distribution
function implies the {\em absence of a quasiparticle pole} in the
one--particle Green's function. Instead one finds in the spectral function
$A(k,\omega)$ a continuum above a threshold $u(k-k_F)$, with a singularity
close to the threshold: $A(k,\omega) \propto (\omega -
u(k-k_F))^{\delta-1}$, and there is also some (non--divergent) spectral
weight at negative energies \cite{lut_bos,voit_spectral,meden_spectral}.

The coefficient $K$ also determines the
long-distance decay of all other correlation functions of the system:
in the present context the most interesting correlations are those involving
charge density or pairing fluctuations.
Using the representation (\ref{singlepsi0}) the corresponding operators
can be written as
\begin{eqnarray}
\label{ocdw0}
O_{CDW}(x) & = & \psi^\dagger_-(x) \psi_+(x) =
\lim_{\alpha \rightarrow 0}
\frac{e^{2ik_Fx}}{\pi\alpha} e^{-2i \phi(x)} \virg \\
\label{osc0}
O_{SC}(x) & = & \psi_-(x) \psi_+(x) =
\lim_{\alpha \rightarrow 0}
\frac{1}{\pi\alpha} e^{-2i \theta(x)}   \point
\end{eqnarray}
Similar relations can also be found for  other operators.
The CDW and pairing correlation functions now are easily calculated
as a function of $x$, $t$, and temperature  \cite{emery_revue_1d}.
I here only give the leading asymptotic behavior at zero temperature:
\begin{eqnarray}
\nonumber
\langle O^\dagger_{CDW}(x,t) O_{CDW}(0) \rangle & = & C_1
(x^2 - u^2 t^2)^{-K}  \virg \\
\langle O^\dagger_{SC}(x,t) O_{SC}(0) \rangle & = & C_2
(x^2 - u^2 t^2)^{-1/K}
 \virg
\end{eqnarray}
with interaction--dependent constants $C_{1,2}$.
The corresponding susceptibilities (i.e. the Fourier transforms of the
above correlation functions) behave at low temperatures as
\begin{equation}
\chi_{CDW}(T) \approx T^{2K -2 }  \virg
\chi_{SC}(T) \approx T^{2/K -2 }   \virg
\end{equation}
i.e. for $K < 1$ charge density fluctuations at
$2 k_F$ are enhanced and
diverge at low temperatures, whereas for $K >1$
pairing fluctuations dominate. The remarkable fact in all the above results
is that there is only {\em one coefficient}, $K$, which determines all
the asymptotic power laws.

To summarize the spinless case, we have found striking differences
with
the usual Fermi liquid picture: there are no quasiparticle poles in the
single--particle Green's function, but rather power law singularities,
with interaction--dependent exponents. Similar power laws appear in all
types of correlation functions. However, all exponents are related to
one parameter $K$, which contains all the interaction dependence.
This type of behavior has
been called {\bf Luttinger liquid} by Haldane \cite{haldane_bosonisation}.

\label{spinhalfsec}
\subsection{Spin--1/2 fermions}
In the case of spin--1/2 fermions, all the fermion operators acquire an
additional spin index $s$. Following the same logic as above, the kinetic
energy then takes the form
\begin{eqnarray}
\nonumber
H_0 & = & v_F \sum_{k,s} \{(k-k_F) a_{k,s}^\dagger a_{k,s} +
 (-k-k_F) b_{k,s}^\dagger b_{k,s} \} \\
\label{h0s}
&=& \frac{2 \pi v_F}{L} \sum_{q>0,s}
[\rho_{+,s}(q) \rho_{+,s}(-q) + \rho_{-,s}(-q) \rho_{-,s}(q)]
\virg
\end{eqnarray}
where density operators for spin projections $s=\uparrow,\downarrow$ have
been introduced:
\begin{equation}
\rho_{+,s}(q) = \sum_k a^\dagger_{k+q,s} a_{k,s} \;\; , \;\;
\rho_{-,s}(q) = \sum_k b^\dagger_{k+q,s} b_{k,s} \;\; \point
\end{equation}
There are now two types of interaction. First, the ``backward scattering''
$(k_F,s;-k_F,t)$ $ \rightarrow $ $ (-k_F,s;k_F,t)$ which for $s \neq t$
cannot
be re--written as an effective forward scattering (contrary to the spinless
case). The corresponding Hamiltonian is
\begin{equation}
H_{int,1}  =  \frac{1}{L} \sum_{kpqst}
g_1 a_{k,s}^\dagger b_{p,t}^\dagger a_{p+2k_F+q,t} b_{k-2k_F-q,s}
\point
\end{equation}
And, of course, there is also the forward scattering, of a form similar to
the spinless case
\begin{eqnarray}
\label{hint2}
H_{int,2}  & = & \frac{1}{2L} \sum_{qst} \{2
g_2(q) \rho_{+,s}(q) \rho_{-,t}(-q) \\
& & + g_4(q) [\rho_{+,s}(q) \rho_{+,t}(-q) + \rho_{-,s}(-q) \rho_{-,t}(q)]\}
\point
\end{eqnarray}

To go to the bosonic description, one introduces $\phi$ and $\Pi$ fields for
the two spin projections separately, and then transforms to charge and spin
bosons
via $\phi_{\rho,\sigma}= (\phi_\uparrow \pm \phi_\downarrow)/\sqrt2$,
$\Pi_{\rho,\sigma}= (\Pi_\uparrow \pm \Pi_\downarrow)/\sqrt2$.
The operators $\phi_\nu$ and $\Pi_\nu$ obey Bose--like
commutation relations:
$$[\phi_\nu(x),\Pi_{\mu}(y)] = i\delta_{\nu\mu}\delta (x-y) \virg $$
and single fermion operators can be written in a form analogous to
(\ref{singlepsi0}):
\begin{equation} \label{singlepsi}
\psi_{\pm,s}(x) = \lim_{\alpha \rightarrow 0} \frac1{\sqrt{2\pi
\alpha}} \exp \left[
{\pm i k_F x}
-i(\pm (\phi_\rho + s \phi_\sigma) - (\theta_\rho+s\theta_\sigma))
/\sqrt{2} \right]
\virg
\end{equation}
where $\theta_\nu(x) = \pi \int^x \Pi_\nu(x') dx'$.

The full Hamiltonian $H = H_0 + H_{int,1} + H_{int,2}$ then takes the form
\begin{equation} \label{hbos}
H = H_\rho + H_\sigma + \frac{2g_1}{(2\pi \alpha )^2}
\int dx \cos (\sqrt8 \phi_\sigma ) \point
\end{equation}
Here $\alpha$ is a short-distance cutoff,
and for $\nu = \rho , \sigma $
\begin{equation} \label{hnu}
 H_\nu = \int dx \left[ \frac{\pi u_\nu K_\nu}{2} \Pi_\nu^2
       + \frac{u_\nu}{2\pi K_\nu} (\partial_x \phi_\nu )^2 \right] \virg
\end{equation}
with
\begin{equation}
\label{uks}
u_\nu = [(v_F+g_{4,\nu}/\pi)^2 - g_\nu^2/(2\pi)^2]^{1/2} \virg \;
K_\nu = \left[ \frac{2\pi v_F +2g_{4,\nu}+ g_\nu}{2\pi v_F +2g_{4,\nu}-
g_\nu}\right]^{1/2} \virg
\end{equation}
and $g_\rho = g_1 -2g_2$, $g_\sigma = g_1$, $g_{4,\rho} = g_4$,
$g_{4,\sigma} = 0$. For a noninteracting system
one thus has $u_\nu = v_F$ (charge and spin velocities equal!) and $K_\nu
=1$. For $g_1=0$,
(\ref{hbos}) describes independent long-wavelength oscillations of the
charge and spin density, with linear dispersion relation $\omega_\nu(k)
= u_\nu |k|$, and the system is conducting. As in the spinless case,
there are no single--particle or single particle--hole pair excited states.
This model (no backscattering), usually called the Tomonaga--Luttinger model,
is the one to which the bosonization method was originally applied
\cite{tomonaga_model,luttinger_model,mattis_lieb_bos}.

For $g_1 \neq 0$ the cosine term has to be treated perturbatively.
A straightforward calculation gives the renormalization group
equation
\begin{equation}
\label{rg1}
 \frac{d}{dl}g_1(l) = \frac{1}{\pi v_F} g_1(l)^2 \virg
 \frac{d}{dl}g_2(l) = \frac{1}{2 \pi v_F} g_1(l)^2 \virg
\end{equation}
where  the renormalized cutoff $E_c$ is related to the bare cutoff $E_0$ via
$E_c = E_0 \exp(l)$. Thus,
 for repulsive interactions ($g_1 >
0$), $g_1$ is renormalized to zero in the long-wavelength limit, and
at the fixed point one has $K_\sigma^* = 1$. The three remaining
parameters in (\ref{hbos}) then completely determine the long-distance
and low--energy properties of the system.

It should be emphasized that (\ref{hbos}) has been derived here for
fermions with linear energy--momentum relation. For more general (e.g.
lattice) models, there are additional operators arising from band
curvature and the absence of high--energy single--particle states.
One can however show that all these effects are, at least for not very
strong interaction, irrelevant in the renormalization group sense, i.e. they
do not affect the low--energy physics. Thus, {\em(\ref{hbos})
is still the correct effective Hamiltonian for low--energy excitations}.

The Hamiltonian (\ref{hbos}) also provides an explanation for the physics of
the case of negative $g_1$, where the renormalization group scales to strong
coupling (eq.(\ref{rg1})). In fact, if $|g_1|$ is large in
(\ref{hbos}), it is quite clear that the elementary excitations of
$\phi_\sigma$ will be small oscillations around one of the minima of the
$\cos$ term, or possibly soliton--like objects where $\phi_\sigma$ goes from
one of the minima to the other. Both types of excitations have a gap, i.e.
for $g_1 < 0$ one has a {\em gap in the spin excitation spectrum}, whereas
the charge excitations remain massless. This can actually investigated in
more detail in an exactly solvable case \cite{luther_exact}.

\subsubsection{Spin--charge separation}
One of the more spectacular consequences of the Hamiltonian (\ref{hbos})
is the complete separation of the dynamics of the spin and charge
degrees of freedom. For example, in general one has $ u_\sigma \neq
u_\rho$, i.e. the charge and spin oscillations propagate with
different velocities. Only in a noninteracting system or if some
accidental degeneracy occurs does one
have $u_\sigma = u_\rho = v_F$. To make the meaning of this fact more
transparent, let us create an extra particle in the ground state, at
$t=0$ and spatial coordinate $x_0$. The charge and spin densities then
are easily found, using
 $\rho(x) = -(\sqrt2/\pi) \partial \phi_\rho/\partial x$
 (note that $\rho(x)$ is the deviation of the
density from its average value) and
 $\sigma_z(x) = -(\sqrt2/\pi) \partial \phi_\sigma / \partial x$
:
\begin{eqnarray}
\nonumber
\langle 0 | \psi_+(x_0) \rho(x) \psi^\dagger_+(x_0) | 0 \rangle & = &
\delta(x-x_0) \virg \\
\langle 0 | \psi_+(x_0) \sigma_z(x) \psi^\dagger_+(x_0) | 0 \rangle & =
& \delta(x-x_0) \point
\end{eqnarray}
Now, consider the time development of the charge and spin
distributions. The time--dependence of the charge and spin density
operators is easily obtained from  (\ref{hbos}) (using the fixed point
value $g_1 = 0$), and one obtains
\begin{eqnarray}
\nonumber
\langle 0 | \psi_+(x_0) \rho(x,t) \psi^\dagger_+(x_0) | 0 \rangle & = &
\delta(x-x_0-u_\rho t) \virg \\
\langle 0 | \psi_+(x_0) \sigma_z(x,t) \psi^\dagger_+(x_0) | 0 \rangle &
= & \delta(x-x_0-u_\sigma t) \point
\end{eqnarray}
Because in general $u_\sigma \neq u_\rho$, after some time charge and
spin will be localized at completely different points in space, i.e.
{\em charge and spin have separated completely}. A interpretation of
this surprising phenomenon in terms of the Hubbard model will be given
in sec.(\ref{hubsec}).

Here a linear energy--momentum relation has been
assumed for the electrons, and consequently the shape of the charge
and spin distributions is time--independent. If the energy--momentum
relation has some curvature (as is necessarily the case in lattice
systems) the distributions will widen with time. However this widening
is proportional to $\sqrt{t}$, and therefore much smaller than the
distance
between charge and spin. Thus, the qualitative picture of spin-charge
separation is unchanged.

\subsubsection{Physical properties}
The simple form of the Hamiltonian (\ref{hbos}) at the fixed point
$g_1^*=0$ makes the calculation of physical properties rather
straightforward. The specific heat now is determined both by the charge and
spin modes, and consequently the specific heat coefficient $\gamma$ is given
by
\begin{equation}       \label{gamma}
\gamma/\gamma_0 = \frac12 (v_F/u_\rho + v_F / u_\sigma) \point
\end{equation}
Here $\gamma_0$ is the specific heat coefficient of noninteracting
electrons of Fermi velocity $v_F$.

The spin susceptibility and the compressibility are equally easy to
obtain. Note that in
(\ref{hbos}) the coefficient $u_\sigma / K_\sigma$ determines the energy
necessary to create a nonzero spin polarization, and, as in the spinless
case, $u_\rho
/ K_\rho$ fixes the energy needed to change the particle density. Given
the fixed point value $K_\sigma^*=1$, one finds
\begin{equation}  \label{chi}
\chi / \chi_0 = v_F/u_\sigma \virg \quad \kappa / \kappa_0 = v_F K_\rho
/ u_\rho \virg
\end{equation}
where $\chi_0$ and $\kappa_0$ are the susceptibility and
compressibility of the noninteracting
case. From eqs.(\ref{gamma}) and (\ref{chi}) the Wilson ratio is
\begin{equation}   \label{rw}
R_W = \frac{\chi }{ \gamma} \frac{\gamma_0}{\chi_0} =
\frac{2 u_\rho}{u_\rho +u_\sigma } \point
\end{equation}

The quantity $\Pi_\rho(x)$ is proportional to the current density.
As before, the Hamiltonian commutes with the total current, one thus has
\begin{equation}  \label{sig0}
\sigma (\omega) = 2 K_\rho u_\rho \delta (\omega ) +
\sigma_{reg}(\omega) \virg
\end{equation}
i.e. the product $K_\rho u_\rho $ determines the weight of the dc peak
in the conductivity. The regular part of the conductivity  in general
varies as $\omega^3$ at low frequencies \cite{giamarchi_millis}.

The above properties, linear specific heat, finite spin susceptibility,
and dc conductivity are those of an ordinary Fermi liquid, the
coefficients $u_\rho, u_\sigma$, and $K_\rho$ determining
renormalizations with respect to noninteracting quantities. However, the
present system is {\em not a Fermi liquid}. This is in fact already
obvious from the preceding discussion on charge--spin separation,
and can be made more precise considering
the single--particle Green's function. Using the
representation (\ref{singlepsi}) of fermion operators one  finds
(at the fixed point $g_1 = 0$)
\begin{eqnarray}
\nonumber
G^R(x,t) & = & -i \theta(t) \langle
[\psi_{+,s}(x,t),\psi_{+,s}^\dagger(0,0)]_+ \rangle \\
\label{gs}
& = & -\frac{\theta (t)}{\pi} e^{ik_Fx} {\rm Re}
\left\{ \frac{1}{\sqrt{(u_\rho t -x) (u_\sigma t -x)}}
  \left[\frac{\alpha^2}{(\alpha+iu_\rho t)^2 + x^2} \right]^{\delta/2}
   \right\}
\point
\end{eqnarray}
Note that this expression factorizes into a spin and a charge contribution
which propagate with different velocities.
Fourier transforming (\ref{gs})  gives
 the momentum distribution function in the vicinity of $k_F$:
\begin{equation}
n_k \approx n_{k_F} -  {\rm const.} \,
 {\rm sign}(k-k_F) |k-k_F|^\delta \;\; \virg
\end{equation}
and for the single-particle density of states (i.e. the
momentum--integrated spectral density) one finds:
\begin{equation}
\label{spec}
N(\omega ) \approx |\omega |^\delta \point
\end{equation}
In both cases $\delta=(K_\rho + 1/K_\rho -2)/4$. Note that for any
$K_\rho \neq 1$, i.e. {\em for any nonvanishing interaction}, the
momentum distribution function and the density of
states have power--law singularities at the Fermi level, with a
vanishing single particle density of states at $E_F$. This
behavior is obviously quite different from a standard Fermi liquid
which would have a finite density of states and a step--like singularity
in $n_k$. The absence of a step at $k_F$ in the momentum distribution
function implies the {\em absence of a quasiparticle pole} in the
one--particle
Green's function. In fact, a direct calculation of the spectral function
$A(k,\omega)$
from (\ref{gs}) \cite{voit_spectral,meden_spectral} shows that the
usual quasiparticle pole is replaced by
a continuum, with a lower threshold at $\min(u_\nu)(k-k_F)$ and branch cut
singularities at $\omega = u_\rho p$ and $\omega = u_\sigma p$:
\begin{equation}
A(k,\omega) \approx (\omega - u_\sigma (k-k_F))^{\delta-1/2} \virg
|\omega - u_\rho (k-k_F)|^{(\delta-1)/2} \quad (u_\rho > u_\sigma)
\virg
\end{equation}
\begin{equation}
A(k,\omega) \approx (\omega - u_\rho (k-k_F))^{(\delta-1)/2} \virg
|\omega - u_\sigma (k-k_F)|^{\delta-1/2} \quad (u_\rho < u_\sigma)
\point
\end{equation}

The coefficient $K_\rho$ also determines the
long-distance decay of all other correlation functions of the system:
Using the representation (\ref{singlepsi}) the charge and spin density
operators at $2k_F$ are
\begin{eqnarray}
\label{ocdw}
O_{CDW}(x) & = & \sum_s \psi_{-,s}^\dagger(x) \psi_{+,s}(x) =
\lim_{\alpha \rightarrow 0}
\frac{e^{2ik_Fx}}{\pi\alpha} e^{-i\sqrt2 \phi_\rho(x)}
\cos [\sqrt2 \phi_\sigma(x)] \virg \\
O_{SDW_x}(x) & = & \sum_{s} \psi_{-,s}^\dagger (x)
\psi_{+,-s} (x)
= \lim_{\alpha \rightarrow 0}
\frac{e^{2ik_Fx}}{\pi\alpha} e^{-i\sqrt2 \phi_\rho(x)}
\cos [\sqrt2 \theta_\sigma(x)] \point
\end{eqnarray}
Similar relations are
also found for other operators. It is important to note
here that all these operators decompose into a product of one factor
depending on the charge variable only by another factor depending only on
the spin field. Using
the Hamiltonian (\ref{hbos}) at the fixed point $g_1^* = 0$ one finds for
example for
the charge and spin correlation functions\footnote{The time- and temperature
dependence is also easily obtained, see
\cite{emery_revue_1d}.}
\begin{eqnarray} \nonumber
\langle n(x) n(0) \rangle & = & K_\rho/(\pi x)^2 + A_1 \cos (2k_Fx)
x^{-1-K_\rho} \ln^{-3/2}(x)  \\
\label{nn}
& & \mbox{} + A_2 \cos (4k_Fx) x^{-4K_\rho} + \ldots \virg \\
\label{ss}
\langle \ve{S}(x) \cdot \ve{S}(0) \rangle & = & 1/(\pi x)^2 + B_1 \cos
(2k_Fx)
x^{-1-K_\rho} \ln^{1/2}(x) + \ldots \virg
\end{eqnarray}
with model dependent constants $A_i,B_i$. The ellipses in (\ref{nn})
and (\ref{ss}) indicate higher harmonics of $\cos (2k_F x)$ which are
present but decay faster than the terms shown here.
Similarly, correlation functions for singlet (SS) and triplet (TS)
superconducting pairing are
\begin{eqnarray}
\nonumber
\langle O^\dagger_{SS}(x) O_{SS}(0) \rangle & = & C x^{-1-1/K_\rho}
\ln^{-3/2}(x) + \ldots \virg \\
\langle O^\dagger_{TS_\alpha}(x) O_{TS_\alpha}(0) \rangle & = & D
x^{-1-1/K_\rho} \ln^{1/2}(x) + \dots \point
\end{eqnarray}
The logarithmic corrections in these functions \cite{finkelstein_logs} have
been studied in detail recently
\cite{voit_logs,giamarchi_logs,affleck_log_corr,singh_logs}.
The corresponding susceptibilities (i.e. the Fourier transforms of the
above correlation functions) behave at low temperatures as
\begin{equation}
\chi_{CDW}(T) \approx T^{K_\rho -1 } |\ln(T)|^{-3/2} \virg
\chi_{SDW}(T) \approx T^{K_\rho -1 } |\ln(T)|^{1/2}  \virg
\end{equation}
\begin{equation}
\chi_{SS}(T) \approx T^{1/K_\rho -1 } |\ln(T)|^{-3/2} \virg
\chi_{TS}(T) \approx T^{1/K_\rho -1 } |\ln(T)|^{1/2}  \virg
\end{equation}
i.e. for $K_\rho < 1$ (spin or charge) density fluctuations at
$2 k_F$ are enhanced and
diverge at low temperatures, whereas for $K_\rho >1$
pairing fluctuations dominate. These correlation functions with their power
law variations actually determine experimentally accessible quantities:
the $2k_F$ and $4k_F$ charge correlations lead to X--ray
scattering intensities $I_{2k_F} \approx T^{K_\rho}$, $I_{4k_F} \approx
T^{4K_\rho-1}$, and similarly the NMR relaxation rate due to $2k_F$ spin
fluctuations varies as $1/T_1 \approx T^{K_\rho}$. The remarkable fact in
all the above results
is that there is only {\em one coefficient}, $K_\rho$, which determines all
the asymptotic power laws.

Correlation functions in the spin--1/2 case (``spin--1/2 Luttinger
liquid'') share one important
property with spinless fermions: they have power--law behavior, with
interaction--dependent powers determined by one coefficient, $K_\rho$.
However, the phenomenon of spin--charge separation adds some additional
features in this case and has spectacular consequences both for
thermodynamical and spectral properties.

\sectio{Umklapp scattering and metal--insulator transitions}
\label{umsec}
\subsection{Half--filling}
\label{halfsec}
In the model discussed in the preceding section, total momentum was
always a conserved quantity, and consequently these models are metallic with
an infinite dc conductivity. However, in a half--filled band one has
$k_F = \pi/2$. Then
 umklapp scattering, transferring
two particles from $-k_F$ to $k_F$, involves momentum transfer $4k_F = 2\pi$
which is a reciprocal lattice vector. These processes are thus allowed and
lead to an extra term
\begin{equation}
\label{hint3}
H_{int,3}  =  \frac{1}{L} \sum_{kpqs}
g_3 (a_{k,s}^\dagger a_{p,-s}^\dagger b_{p-2k_F+q,-s} b_{k-2k_F-q,s} +
h.c.)
\point
\end{equation}
in the Hamiltonian. Note that because of the Pauli principle, only
scattering of two particles of opposite spin is allowed if, as usually
is assumed, $g_3$ is momentum--independent. In the boson representation,
this term
leads to an additional interaction for the charge degrees of freedom:
\begin{equation}
\label{g3}
H_{int,3}  =
\frac{2g_3}{(2\pi \alpha )^2} \int dx \cos (\sqrt8 \phi_\rho )
\point
\end{equation}
Similarly to the $g_1$ term, this term can be handled via a renormalization
group calculation. The equations are
\begin{equation}
 \frac{d}{dl}g_3(l) = \frac{1}{\pi v_F} (g_1(l)-2g_2(l))g_3(l) \virg
 \frac{d}{dl}(g_1(l)-2g_2(l)) = \frac{1}{ \pi v_F} g_3(l)^2 \point
\end{equation}
These in fact are the well--known Kosterlitz--Thouless equations. There are
two regimes: (i) for $|g_3| \leq g_1 - 2g_2$ $g_3$ scales to zero, and
$g_1 -2g_2$ to the fixed point value $(g_1 - 2g_2)^* = [(g_1 - 2g_2)^2 -
g_3^2]^{1/2}$. Consequently, the
charge excitation spectrum remains massless, but there are corrections of
order $g_3^2$ to $K_\rho$: in eq (\ref{uks}) for $K_\rho$ one replaces
$g_1 - 2g_2 \rightarrow (g_1 - 2g_2)^*$, but $u_\rho$ is
unrenormalized at this order. Of course,
in this case the system remains a metal. (ii) for $|g_3| > g_1 - 2g_2$
$g_3$ scales towards strong coupling. From
eq.(\ref{g3}) one then expects $\phi_\rho$ to be essential fixed to one
 of the minima of the cosine potential. This gives rise to a gap in the
 charge excitation spectrum
\cite{emery_etal_umklapp,solyom_revue_1d}. The ground state then is
insulating.

In a particular but illuminating case, the umklapp problem can be solved
exactly \cite{emery_etal_umklapp}: after the unitary transformation
$\phi_\rho \rightarrow \phi_\rho/\sqrt2$, $\Pi_\rho \rightarrow
\sqrt2 \Pi_\rho$, the total Hamiltonian for the charge degrees of freedom
becomes
\begin{equation}
\label{h30}
 H_\rho = \int dx \left[ \pi u_\rho K_\rho \Pi_\rho^2
       + \frac{u_\rho}{4\pi K_\rho} (\partial_x \phi_\rho )^2 \right]
+\frac{2g_3}{(2\pi \alpha )^2} \int dx \cos (2 \phi_\rho )
\point
\end{equation}
Using the transformations discussed in sec. \ref{wcsec}, this problem can
be transformed into a spinless fermion model. In particular, for
$K_\rho  = 1/2$, the corresponding spinless model has $K=1$ (cf. eq.
(\ref{hbos0})), i.e. $g_2 = 0$. Using eq.(\ref{ocdw0}) $H_\rho$ then
transforms into
\begin{equation}
\label{h31}
H_\rho = v_F \sum_{k} \{(k-k_F) a_{k}^\dagger a_{k} +
 (-k-k_F) b_{k}^\dagger b_{k} \}
 + \frac{g_3}{2\pi\alpha} \sum_k (a_{k}^\dagger b_{k-2k_F}+b_{k}^\dagger
a_{k+2k_F}) \point
\end{equation}
This form is easily diagonalized, and one finds an excitation
spectrum for the spinless fermions of the form (noting that for
half--filling $k_F = \pi/2$)
\begin{equation}
\label{egap}
E_k = \pm [v_F^2 (k \pm \pi/2)^2 + \Delta^2]^{1/2} \virg
\end{equation}
with a gap
\begin{equation}
\Delta = g_3/(2 \pi \alpha) \point
\end{equation}
In the ground state all negative energy states are filled, all
positive energy states are empty, and because of the gap the system then is
an insulator. Note that the fermion operators in (\ref{h31}) in fact
create solitons in the $\phi_\rho$ field, without affecting $\phi_\sigma$.
At $K_\rho \neq 1/2$, the physical picture is not changed qualitatively, in
particular there is still a gap in the charge excitation spectrum, however,
the functional dependence is changed: in general, one has $ \Delta \propto
g_3^\nu$, with $\nu = 1/(2-2K_\rho)$.

The spin field $\phi_\sigma$ is completely unaffected by the
metal--insulator transition. Consequently, even the insulating state has a
Pauli--like susceptibility, and long--range spin correlations of the type
(\ref{ss}), with $K_\rho = 0$. This behavior is that of the
one--dimensional antiferromagnetic Heisenberg model, which can be
considered as a case of electrons localized on individual atoms. In the
present case, the gap is typically much smaller than the bandwidth, and
the localization length is therefore rather big. Nevertheless, the
qualitative behavior of the correlations is the same.

We thus here have an example of a metal--insulator transition, occuring
at constant particle density (half--filling) as a function of the
interaction parameters. Approaching the transition from the metallic
side, one has $K_\rho \rightarrow 1$, $ u_\rho \rightarrow const.$, and
thus in particular the Drude weight of the conductivity remains finite
and jumps discontinuously to zero as the transition line is crossed.
This jump is nothing but the usual jump of the stiffness
parameter (or ``superfluid density'') of the Kosterlitz--Thouless
transition. On
the insulating side, the gap opens exponentially as a function of
interaction strength, again the typical Kosterlitz--Thouless behavior.

One clearly would like to understand what happens if one is close to but not
exactly half--filled. One thus adds a chemical potential term to the
original model to obtain
\begin{equation}
H = H_0 + \sum_{i=1}^3 H_{int,i} - \mu \sum_{k,s}(a_{k,s}^\dagger a_{k,s}
+ b_{k,s}^\dagger b_{k,s}) \point
\end{equation}
In the bosonic description, the chemical potential term gives rise to a term
proportional to $\partial \phi_\rho/\partial x$, which in turn, in the
spinless fermion language of eq.(\ref{h31}) is equivalent to a chemical
potential. For $K_\rho=1/2$, the solution is still straightforward: as long
as $|\mu| < \Delta$, the ground state is unchanged, but for
$\mu > \Delta$, positive energy states start to be occupied (and similarly,
for $\mu <- \Delta$, negative energy states are emptied). The number of
extra carriers (i.e. the deviation of the carrier density from
half--filling) varies
as $\delta n \propto \sqrt{|\mu | - \Delta}$. The system is now metallic,
and from eq.(\ref{egap}) one obtains the effective velocity of the charge
modes as
\begin{equation}
u_\rho = \frac{v_F^2 |k_F -\pi/2|}{[v_F^2 (k_F - \pi/2)^2 +
\Delta^2]^{1/2}} \end{equation}
In particular, $u_\rho$ vanishes linearly as half--filling is approaches.
Thus, the Drude weight of the conductivity vanishes linearly
as $n \rightarrow 1$.

If $K_\rho \neq 1/2$ in eq.(\ref{h30}) there are additional interaction
terms between the spinless
fermions of eq.(\ref{h31}). However, in the vicinity of half--filling, these
interactions can be eliminated from the problem
\cite{schulz_cic2d,giamarchi_rho}, and one has then
for the effective parameter governing the low--energy physics
$K_{\rho}^* = 1/2$ in all cases, even if the original $K_\rho$ was quite
different from $1/2$.

It should be emphasized here that $K_{\rho}^* = 1/2$ is valid
for $n \rightarrow 1$, but not for $n=1$. In the latter case,
there is a gap in the charge excitation spectrum due to the
umklapp term (\ref{g3}), and the correlations of $\phi_\rho$ become long
ranged, i.e. $K_{\rho}^* = 0$. Close to half--filling, the asymptotic
behavior of the charge part
of correlation functions like (\ref{ss}) is essentially determined by the
motion of the added carriers. Writing the density of carriers  as $\rho
= |1-n|$ one then expects a crossover of the form \cite{schulz_cic2d}
\begin{equation}
\label{cross}
\langle \ve{S}(x) \cdot \ve{S}(0) \rangle \approx \cos (2k_Fx) [1 + (\rho
x)^2 ]^{-K_\rho/2} x^{-1} \ln^{1/2}(x)
\end{equation}
for the $2k_F$ part of the spin correlation function, and similarly for
other correlation functions. Clearly, only for $x \gg 1/\rho$ are the
asymptotic power laws valid, whereas at intermediate distances $ 1 \ll x \ll
1/\rho$ one has effectively $K_\rho = 0$. The form (\ref{cross})
provides a smooth crossover as $n \rightarrow 1$.

An interesting question is the sign of the charge carriers, especially
close to the metal--insulator transition. The standard way to determine
this, the sign of the Hall constant, is useless in a one--dimensional
system. As an alternative, the thermopower can be used which is negative
(positive) for electron (hole) conduction. In general, calculation of
the thermopower is a nontrivial task, as the curvature of the bands
plays an important role, and the approximate form of the Hamiltonian
(\ref{hbos}) is therefore insufficient. Moreover, both charge and spin
entropies can play a role. However, close to the metal--insulator
transition $u_\rho \ll u_\sigma$, and therefore the entropy of the
charge degrees of freedom is much bigger than the spin entropy. As discussed
above, the charge part of the Hamiltonian can be transformed
into a model of massive fermions, with energy--momentum relation
given by (\ref{egap}). At half--filling all
negative energy states are filled, all positive energy states are empty.
Doping with a concentration $n^*$ of holes, some of the negative energy
states become empty and only states with $|k| > k_F^* \propto n^*$ are
filled. Because of the vanishing interaction, a standard formula for the
thermopower can be used  \cite{chaikin_thermopower} and gives
\begin{equation}
S = \frac{\pi^2 k_B^2 T}{6|e|} \frac{\Delta^2}{v^2 (k^*_F)^2
 (v^2 (k^*_F)^2 + \Delta^2)^{1/2}} \virg
\end{equation}
i.e. {\em approaching the metal--insulator transition from $n < 1$, the
thermopower is hole--like}, whereas obviously far from the transition
($n \ll 1$) it is electron--like. The exactly opposite behavior can be
found for $n > 1$.

At zero temperature and away from half--filling, one has an infinite dc
conductivity in this model. However, at finite temperature there is some
probability to excite a carrier into a momentum state that makes
umklapp scattering possible. One then expects a conductivity increasing
exponentially at low temperature \cite{giamarchi_rho,giamarchi_millis}.

For spinless fermions, a term like (\ref{hint3}) cannot act, because of
the Pauli principle. However, a term like
\begin{equation}
\int dx \; \{ \psi_+^\dagger(x) [\partial_x \psi_+^\dagger(x)]
\psi_-(x) [\partial_x \psi_-(x)] + h.c. \}
\approx \int dx \; \cos(4 \phi(x))
\end{equation}
does produce umklapp scattering at half--filling and can give rise to a
gap in the charge excitation spectrum if $K \le 1/4$
\cite{haldane_xxzchain,bla_equ,nijs_equivalence}. The metal--insulator
transition has properties analogous to that in the spin--1/2 case and
has been studied in detail by Shankar \cite{shankar_spinless}. A lattice
model having this type of transition is the XXZ spin chain, which via a
Jordan--Wigner transformation can be seen as a model of interacting
spinless fermions.

\subsection{Other commensurabilities}
\label{ocsec}
The umklapp operator (\ref{g3}) is effective at and near half--filling. One
might wonder whether higher--order ``commensurability'', i.e. in a third--
or quarter--filled band, can also lead to insulating states, and at least
intuitively it seems clear that this should be possible. On the other hand,
for such cases, simple two--particle processes cannot be simultaneously
momentum conserving (even modulo a reciprocal lattice vector) and only
involve states near the Fermi energy. However, many--particle processes
{\em induced by bare two--particle interactions do exist:} consider the case
of a third--filled band ($k_F = \pi/3$). In a first step then one has a
(non--umklapp) process of the type $(-\pi/3;-\pi/3) \rightarrow
(-\pi;\pi/3)$, with the particle at $-\pi$ in a high--energy state at the
band edge. In a second step one makes the transition
$(-\pi=\pi;-\pi/3) \rightarrow(\pi/3;\pi/3)$. Umklapp here intervenes due to
the identification of states at $-\pi$ and $\pi$, and the final result is
the transfer of three particles from $-k_F$ to $k_F$. For weak two--particle
interactions with an interaction matrix element $g$ the effective matrix
element $g_u$ for this type of process is of order $g^2/t$ (with bandwidth
$t$),
but for stronger interactions it is hard to estimate. The generalization to
bandfilling $1/m$ (i.e. $2/m$ particles per site) is straightforward:
one transfers
$m$ particles from
$-k_F$ to $k_F$, with matrix element $g_u \approx g^{m+1}/t^m$.

To take the existence of this type of processes into account in the original
continuum description, one adds a term
\begin{equation}
H_u = g_u \int dx \, \{[\psi_-^\dagger(x)]^m [\psi_+(x)]^m + h.c. \} \virg
\end{equation}
where the product of many fermion operators at one site is to be understood
as point--split, and the spin summation is implied. In the bosonic
language, this term translates into
\begin{eqnarray}
H_u & = & g_u \int dx \, \{ O_{CDW}(x)^m + h.c. \}   \\
  &  \approx & g_u \int dx \, \cos^m(\sqrt2 \phi_\sigma(x))
    \cos (\sqrt2 m \phi_\rho(x)) \point
\end{eqnarray}
There are now two physically rather different cases, according to whether
$n$ is even or odd. Consider first the even case. Then the $\cos^m(\sqrt2
\phi_\sigma(x))$ term can be expanded, and the lowest--order (most relevant
in the renormalization group sense) comes from the constant in this
expansion, e.g. the effective $H_u$ is
\begin{equation}
\label{hue}
H_u \approx g_u \int dx \, \cos (\sqrt2 m \phi_\rho(x)) \point
\end{equation}
After a simple unitary transformation rescaling $\phi_\rho$, we have now a
problem formally identical to the half--filled case. At filling exactly
equal to $1/m$ we thus have an insulating or a metallic phase, with the
metallic state stable for $K_\rho \geq 4/m^2$, and $K_\rho^* = 4/m^2,
u_\rho^* = const.$ at the metal--insulator transition. For $m >2$ the
condition for the existence of an insulating state, $K_\rho < 4/m^2$,
usually requires strongly repulsive interactions, and even then can not
always be satisfied: e.g. for the Hubbard model, $K_\rho > 1/2$ even for
infinitely strong repulsion. In the insulating state, $\phi_\rho$ has
only small oscillations around a fixed average value, and thus the
effective $K_\rho$ is zero. We then have long--range charge density
wave order at wavevector $4k_F$, but still the spin--spin correlations
typical of the one--dimensional antiferromagnetic Heisenberg model (cf.
eq. (\ref{ss})). We note that the solitons in this case have a
jump in $\phi_\rho$ of $\sqrt2 \pi/m$, and thus carry charge $2/m$. As in
the half--filled case, the magnetic properties are largely independent on
whether one is in the insulating or metallic state.

Varying the particle density at constant $K_\rho < 4/m^2$, one recovers
behavior similar to the nearly--half--filled case above: as $ n
\rightarrow 2/m$ one has $u_\rho \rightarrow 0$ and $K_\rho \rightarrow
2/m^2$. Of course, for $m = 2$ we recover precisely the results of the
preceding section.

Things are a bit more complicated if $m$ is odd. In this case the most
relevant term in $H_u$ takes the form
\begin{equation}
\label{huo}
H_u \approx g_u \int dx \, \cos (\sqrt2 \phi_\sigma(x))
\cos (\sqrt2 m \phi_\rho(x)) \virg
\end{equation}
e.g. spin and charge degrees of freedom are coupled. The scaling dimension
of this operator is $(1+ m^2 K_\rho)/2$, and consequently this operator is
relevant and produces a gap if $K_\rho < 3/m^2$. The insulating state is
characterized by
a large effective $g_u$, and consequently, from (\ref{huo}) one expects
both $\phi_\rho$ and $\phi_\sigma$ to oscillate around stable equilibrium
positions, i.e. there is a gap in both the charge and the spin excitations,
and the ground state is non--magnetic and one has $2k_F$ CDW ordering
(in contrast to the paramagnetic ground state and $4k_F$ CDW found for
$m$ even),
There are now separate soliton--like
excitations for charge and spin, carrying charge $2/m$ and spin $1/2$,
respectively.

In a particular case, an exact solution of the odd--$m$ problem can be
given: assume $g_1=0$ (which anyway is the fixed--point value for repulsive
interactions). Then, after the unitary transformation $\phi_\rho \rightarrow
\phi_\rho/m$, the total Hamiltonian, e.g. the sum of (\ref{hbos}) and
(\ref{huo}) becomes
\begin{eqnarray}
\nonumber
 H & = & \int dx \left[ \frac{\pi u_\rho K_\rho m^2}{2} \Pi_\rho^2
       + \frac{u_\rho}{2\pi K_\rho m^2} (\partial_x \phi_\rho )^2 \right] \\
\nonumber
&& + \int dx \left[ \frac{\pi u_\sigma}{2} \Pi_\sigma^2
       + \frac{u_\sigma}{2\pi} (\partial_x \phi_\sigma )^2 \right] \\
\label{huo1}
&&+ g_u \int dx \, \cos (\sqrt2 \phi_\sigma(x))
\cos (\sqrt2 \phi_\rho(x)) \virg
\end{eqnarray}
The $g_u$--term now has exactly the form of the $2k_F$ CDW operator
(\ref{ocdw}), i.e. it is bilinear in fermion operators. Moreover, for
$K_\rho=1/m^2$ and $u_\rho = u_\sigma = u$, the first two terms in
(\ref{huo1}) represent free spin-1/2 fermions, e.g. in this case the full
Hamiltonian is in fact the spin-1/2 version of (\ref{h31}):
\begin{eqnarray}
\nonumber
H & = & v_F \sum_{k,s} \{(k-k_F) a_{k,s}^\dagger a_{k,s} +
 (-k-k_F) b_{k,s}^\dagger b_{k,s} \} \\
\label{h31s}
&& + g_u \sum_{k,s} (a_{k,s}^\dagger b_{k-2k_F,s}+b_{k,s}^\dagger
a_{k+2k_F,s}) \point
\end{eqnarray}
This form is of course easily diagonalized, and,
as expected one has a gap both in the charge and in the spin
excitations. In this particular case these gaps are equal, but
generally, e.g. for $K_\rho \neq 1/m^2$, this will not bee the case.

If one is away from filling $1/m$, an additional chemical potential term
appears in (\ref{h31s}), and either states above the gap get filled or
states below the gap become empty. This is very similar to the
nearly--half--filled case. Because we have effectively noninteracting
electrons in (\ref{h31s}), one still has $K_\rho = 1/m^2$, and $u_\rho
\propto |n -2/m|$. However, and {\em contrary to the case of $m$ even},
now the spin velocity shows critical behavior: $u_\sigma \propto |n -
2/m|$, leading in particular to a diverging spin susceptibility.

This behavior is however not generic: if $K_\rho$ deviates only slightly
from $1/m^2$, or one of the other solvability conditions is not
satisfied, there are interaction terms in addition to the free fermion
Hamiltonian (\ref{h31s}). Close to filling $1/m$ there are very few
extra fermions, compared to $n = 2/m$. However, and
contrary to the spinless case relevant for
$m$ even, interactions do have dramatic effects on spin--1/2 fermions
even in the very dilute limit. In fact, as long as the interaction is
not long--ranged, for sufficiently high dilution (more precisely, if the
interparticle distance is bigger than the scattering length), the
precise form of the interaction matrix element is expected to be
unimportant. One can then take over exact results available for the one
dimensional Hubbard model in the dilute limit (see sec. \ref{lutt}).
In particular, if the interactions added to (\ref{h31s}) are repulsive (for
$u_\rho = u_\sigma$ this corresponds to $K_\rho < 1/m^2$), one
has $u_\rho \propto |n-2/m|$, but now $K_\rho^* = 1/(2m^2)$, and in
particular $u_\sigma \propto (n-2/m)^2$, i.e. there is a very strong
divergence of the spin susceptibility. On the other hand, for attractive
extra interactions in (\ref{h31s}), a spin gap opens close to $n=2/m$, and
one then has $u_\rho \propto |n-2/m|$, $K_\rho^* = 1/m^2$.
A summary of the different
types of critical behavior of the Luttinger liquid parameters $K_\rho$ and
$u_{\rho,\sigma}$ and of some derived physical quantities is given in table
\ref{tab1}.
\begin{table}[htb]
\caption[toto]{Critical behavior of the different
metal--insulator transitions considered in this paper. The transition is
approached from the metallic side. For additional explanations on the last
column see the text.}
\label{tab1}
\begin{tabular}{ccccc}
\hline
& \parbox{2.5cm}{$n = 2/m$ \\ $K_\rho \rightarrow K_\rho^*$\\$m$ even}
& \parbox{2.5cm}{$n = 2/m$ \\ $K_\rho \rightarrow K_\rho^*$\\$m$ odd}
& \parbox{2.5cm}{$n \rightarrow 2/m$ \\ $K_\rho = {\rm const.}$\\$m$ even}
& \parbox{2.5cm}{$n \rightarrow 2/m$ \\ $K_\rho = {\rm const.}$\\$m$ odd}
\\
\hline
\hline
$K_\rho^*$ & $4/m^2$ & $3/m^2$ & $2/m^2$ & $1/(2m^2)$\\
$u_\rho$ &  const. & const.& $|n-1/m|$ & $|n-1/m|$\\
$u_\sigma$ &  const. & const.&   const. & $(n-1/m)^2$\\
$\gamma $ &  const. & const.& $1/|n-1/m|$  & $1/|n-1/m|$\\
$\chi $ &  const. & const.& const.  & $1/(n-1/m)^2$\\
$R_W $ &  const. & const.& $0$ & $2$\\
\hline
\end{tabular}
\end{table}

\begin{figure}
\centerline{
\mbox{
\rotate{
\epsfysize=12cm
\epsffile{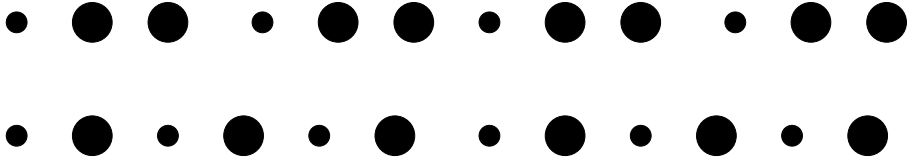}}}}
\caption[f5]{\em Charge arrangement for a system with strong on--site and
nearest--neighbor repulsion for a third--filled (top) and quarter--filled
(bottom) band. Large and small dots are occupied and empty sites,
respectively. Note the alternation of nearest--neighbor distances in the
third--filled case.}
\label{eo}
\end{figure}

The differences between even and odd filling fraction $m$ may a priori
seem surprising. However, there is a simple explanation, illustrated in
figure \ref{eo}. Assume a lattice model with strong finite--range
repulsion. Then, for $m$ even, the ground state will essentially be a
regular sequence of atoms, with one fermion every $m/2$ lattice sites.
The spins of the fermions will interact via an exchange term that is a
generalization of the usual $t^2/U$ term, so that in fact we have a
$4k_F$ charge density wave forming an antiferromagnetic spin chain. This
is exactly the ground state we derived from (\ref{hue}).

On the other hand, for odd $m$ a completely equidistant arrangement is
not possible, as illustrated in fig.\ref{eo} for $m=3$. In fact, a band
filling of $1/m$ corresponds to an average $2/m$ fermions per site.
Consequently, the ground state arrangement is an alternation of short
and long ``bonds'', and one has a $2k_F$ CDW. In addition, there is of
course an alternation of the exchange constants from long to short
bonds, and such an alternation is well--known to introduce a gap in the
spin excitation spectrum. We thus again explain the behavior derived
from (\ref{huo}).

The above considerations can be extended to obtain a lower bound on the
correlation exponent $K_\rho$ in a certain class of models: consider a
lattice model with a finite range interaction of the form
\begin{equation}
\label{hv}
H_{int} = \sum_{i,m\ge 0} V_m n_i n_{i+m} \point
\end{equation}
Let us assume further that $V_m=0$ for $m \ge m_0$, where $m_0$ thus
specifies the range of the interaction ($m_0 = 1$ and $2$ for the Hubbard
and extended Hubbard models discussed below). I will further assume strictly
repulsive interactions, e.g. $V_m > V_{m+1}$. Then, in the strong coupling
limit it is quite clear that the longest interparticle distance in a stable
structure is $m_0$. In particular, the most dilute stable structure has
just one particle every $m_0$ sites, i.e. it has period $m_0$, and
commensurate states with period larger than $m_0$ are not stable
(stability here implies the existence of a gap in the charge
excitation spectrum).
In the continuum limit, a structure with period say $m_0+1$ can be
stabilized by a term $\approx \cos(\sqrt8 (m_0+1) \phi_\rho )$, which has
scaling dimension $ x_{m_0+1} =2(m_0+1)^2 K_\rho$. This term will generate a
stable commensurate structure if it is relevant, i.e. if $x_{m_0+1} \le 2$.
On the other hand, for it {\em not} to create a stable commensurate
structure (as expected), one needs $x_{m_0+1} > 2$. Thus, for purely
repulsive models with interaction of the form (\ref{hv}) and range $m_0$
one has the limit
\begin{equation}
\label{klim}
K_\rho > \frac{1}{(m_0+1)^2} \point
\end{equation}
The assumption we have made here is that quite generally $K_\rho$ decreases
if the interaction term (\ref{hv}) is increased by applying a overall scale
factor that is larger than unity. The limit (\ref{klim}) is satisfied for
the lattice models discussed in the next section.

\sectio{The Hubbard model in one dimension}
\label{hubsec}
\subsection{The Hamiltonian and its symmetries}
The Hubbard model is the prototypical model used for the description of
correlated fermions in a large variety of circumstances, ranging from
high--$\rm T_c$ superconductors to heavy fermion compounds and organic
conductors. In spite of its apparent simplicity, there is still no general
solution, or even a consensus on its fundamental properties. Notable
exceptions are the cases of one and infinite dimensions
\cite{vollhardt_jeru,georges_infdim}. In particular,
in one dimension an exact solution is available. This solution gives
exact energies of the ground state and all the excited states in terms of
the solution of a system of coupled nonlinear equations. On the other hand,
the corresponding wavefunctions have a form so complicated that the explicit
calculation of matrix elements, correlation functions and other physical
quantities has remained impossible so far. In the following sections I shall
describe in some detail the energy spectrum obtained from the exact
solution. Subsequently,
I will show how the knowledge of the energy spectrum can be
combined with the results of the preceding two chapters to obtain a rather
detailed picture of the low--energy properties, in particular of correlation
functions, and of the metal insulator transition.

The Hamiltonian in one dimension has the well--known form
\begin{equation}
H = -t \sum_{i,s} (c_{i,s}^\dagger c_{i+1,s} + c_{i+1,s}^\dagger c_{i,s})
+ U \sum_i n_{i,\uparrow} n_{i,\downarrow} \virg
\end{equation}
where $c_{i,s}$ is the fermion annihilation operator on site $i$ with spin
$s$, $n_{i,s}$ is the corresponding number operator, and the sum is over the
$L$ sites of a one--dimensional chain with periodic boundary conditions.

The model has {\em two} global SU(2) symmetries
\cite{pernici_su2,zhang_su2,schulz_su2}:
the first is the well--known spin rotation invariance, with generators
\begin{equation}
\zeta = \sum_{i=1}^L c_{i,\uparrow}^\dagger c_{i,\downarrow},
\quad \zeta^\dagger
= (\zeta)^\dagger, \quad \zeta_z = \frac12 \sum_{i=1}^L (n_{i,\downarrow}
-n_{i,\uparrow}) \point
\end{equation}
The second type of symmetry is particular to the Hubbard model and relates
sectors of different particle numbers. Its generators are
\begin{equation}                          \label{su2c}
\eta = \sum_{i=1}^L (-1)^i c_{i,\uparrow} c_{i,\downarrow}, \quad
\eta^\dagger
= (\eta)^\dagger, \quad \eta_z = \frac12 \sum_{i=1}^L (n_{i,\downarrow}
+n_{i,\uparrow}) -\frac{L}{2}\point
\end{equation}
The total symmetry thus is $SU(2) \times SU(2) \simeq SO(4)$. One should
notice that more complicated interactions, e.g. involving further
neighbors, will conserve the spin rotation invariance but in general
not the ``charge'' SU(2) invariance (\ref{su2c}). Rather, this second
symmetry will become the standard global $U(1)$ invariance associated with
particle number conservation. It is nevertheless possible to construct
particular types of further--neighbor interactions which do conserve the
full $SU(2) \times SU(2)$ invariance.

\subsection{The exact solution: ground state and excitations}
\label{exsec}
The exact wavefunctions of the one--dimensional Hubbard model for $N$
particles
are superpositions of plane waves characterized by a set $\{ k_j \}$ of
momenta \cite{lieb_hubbard_exact}.
The allowed values of $k_j$ are obtained from the
solution of the coupled set of nonlinear equations
\begin{eqnarray}                      \label{be1}
e^{ik_jL} & = & \prod_{\alpha = 1}^M e\left(\frac{4(\sin k_j -
\lambda_\alpha)}{U} \right) \\
\label{be2}
\prod_{j=1}^N e\left( \frac{4(\lambda_\alpha - \sin k_j)}{U} \right) & = &
- \prod_{\beta=1}^M e \left( \frac{2(\lambda_\alpha - \lambda_\beta)}{U}
\right) \virg
\end{eqnarray}
Here $M$ is the number of down--spin
electrons ($M \le N/2$) and $e(x) = (x+i)/(x-i)$. The $\lambda_\alpha$ are
parameters characterizing the spin dynamics. We note that in general, both
the $k_j$'s and the $\lambda$'s are allowed to be complex. The energy and
momentum of a state are
\begin{equation}
E = -2t \sum_{j=1}^N \cos k_j \quad \virg \quad \quad P = \sum_{j=1}^N k_j
\point
\end{equation}

\subsubsection{Solutions of the Bethe ansatz equations}
The determination of all the solutions of eqs.
(\ref{be1}, \ref{be2}) is not easy. It has recently been shown (under
certain assumptions) that these
equations do indeed give all the ``lowest weight'' (with respect
to $SU(2) \times SU(2)$) eigenstates of the Hubbard model, i.e. all states
satisfying $\eta | \psi \rangle = \zeta | \psi \rangle =0$. The complete
set of eigenstates then is obtained acting repeatedly with
$\eta^\dagger$ or $\zeta^\dagger$ on $|\psi\rangle $
\cite{essler_hubbard_complete}. Here I will limit myself to the ground state
and to the low--lying elementary excitations. These questions have been
investigated in some detail
\cite{shiba_hubbard_exact,coll_excitations,woynarovich_spin}, however, the
finite chain data presented below seem to be quite useful in understanding
the nature of the excitations, and I therefore discuss them in detail.

If both the
$k$'s and the $\lambda$'s are all real, only the phases in (\ref{be1},
\ref{be2}) have to be determined. Taking the logarithm of these equations,
one finds
\begin{eqnarray}
\label{be3} \lefteqn{
L k_j  =  2 \pi I_j + 2\sum_{\alpha=1}^M \arctan [4 (\lambda_\alpha - \sin
k_j)/U]}  \virg & & \\
	    \label{be4} \lefteqn{
2 \sum_{j=1}^N \arctan [ 4 (\lambda_\alpha - \sin k_j)/U]  =
2 \pi J_\alpha + 2 \sum_{\beta =1}^M \arctan [ 2(\lambda_\alpha -
\lambda_\beta)/U] \point}  & &
\end{eqnarray}
The quantum numbers $\{I_j \}$ are all distinct from each other and are
integers if $M$ is even and half--odd integers (HOI, i.e. of the form 1/2,
3/2, \dots ) if $M$ is odd, and are only defined modulo $L$.
Similarly,
the set $\{J_\alpha \}$ are all distinct and are integers if $N-M$ is odd
and HOI if $N-M$ is even. Moreover, there is the restriction
\begin{equation}    \label{restj}
|J_\alpha| < (N-M+1)/2 \point
\end{equation}
Summing (\ref{be3}) over $j$ and (\ref{be4}) over $\alpha$, the total
momentum is found as
\begin{equation}
P = \frac{2 \pi}{L} \left(\sum_{j=1}^N I_j + \sum_{\alpha=1}^M J_\alpha
\right) \point
\end{equation}

\paragraph{Ground state.}
The ground state is nondegenerate only if $N$ is of the form $4\nu + 2$
($\nu$ integer): obviously, if $N$ is odd, the ground state has (at least)
spin 1/2. Further, if $N$ is an integer multiple of 4 the
noninteracting ground state has a sixfold degeneracy, and in the interacting
case the ground state turns out to be a spin triplet. In the following I
shall restrict myself to the case of $N = N_0 = 4 \nu+2$, i.e. the ground
state is
nondegenerate (in the following, $N_0$ will denote the particle number in
the ground state). The ground state then is a singlet, with $M = N_0/2$,
i.e. $M$ is odd. The allowed values of the $J$'s range from $-(N_0/2-1)/2$ to
$(N_0/2-1)/2$. There are exactly $N_0/2$ such integers, i.e. all the $J$'s
are fixed. The $I$'s are consecutive between $-(N_0-1)/2$ and $(N_0-1)/2$,
i.e. \begin{eqnarray}
\label{si1}
\{ I_j \} & = & \{-(N_0-1)/2, \dots , (N_0-1)/2 \} \virg \\
\{ J_\alpha \} & = & \{-(N_0/2-1)/2, \dots , (N_0/2-1)/2 \} \point
 \end{eqnarray}
In the
thermodynamic limit $L \rightarrow \infty$ the distance between consecutive
$k$'s or $\lambda$'s decrease like $1/L$, and one can then find
linear integral
equations for the density of $k$'s and $\lambda$'s on the real axis.
Numerical results for the ground state energy as a function of particle
density and $U$ have been given by Shiba \cite{shiba_hubbard_exact}.

At half--filling, there is a gap in the charge excitations
\cite{lieb_hubbard_exact},
and only spin excitations ( the ``$2k_F$'' triplet and singlet states below)
are gapless. In particular, a finite energy is required to add or take out
a particle. On the other
hand, away from half--filling, both charge and spin excitations are massless.

\begin{figure}
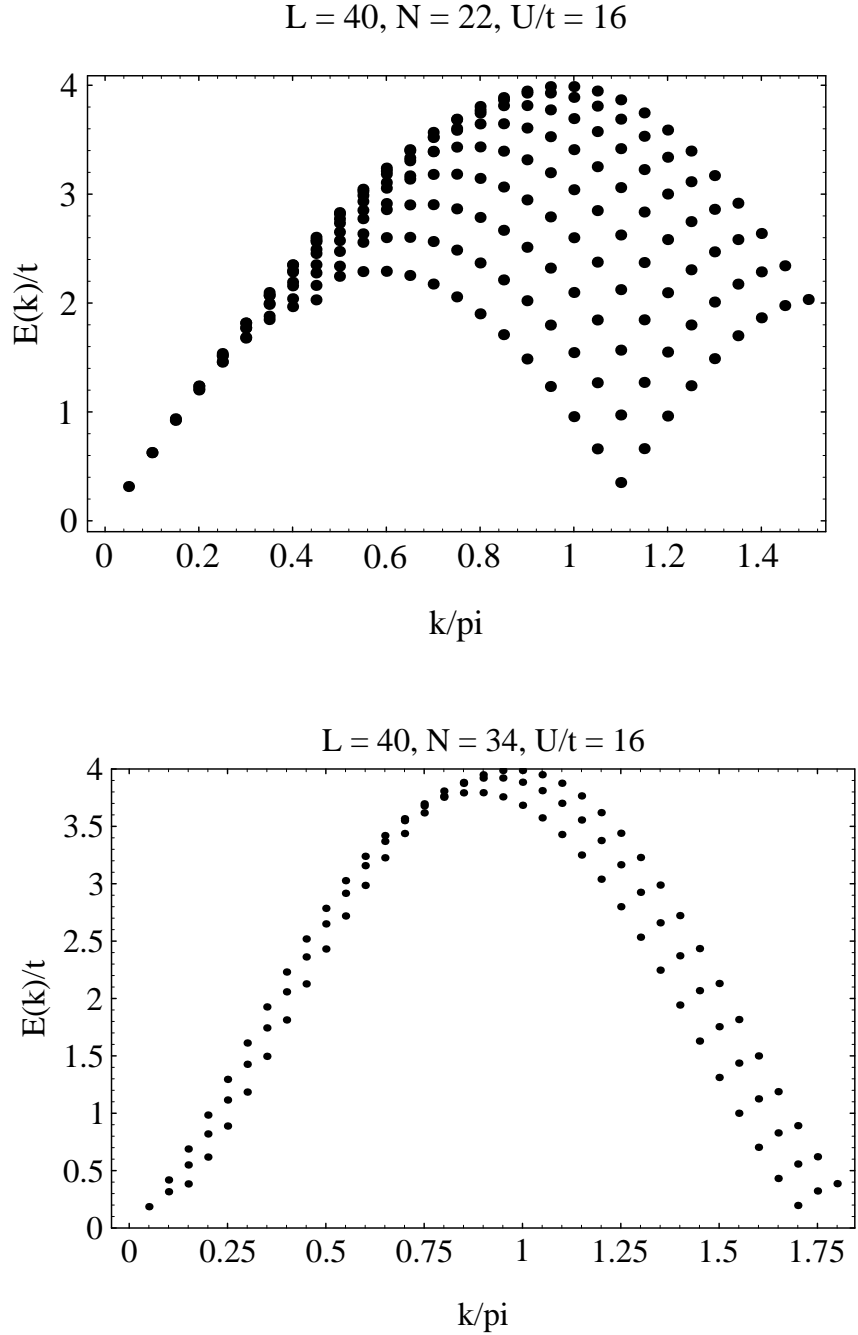

\centerline{
\mbox{
\rotate{
\epsfysize=12cm
\epsffile{/u/heinz/tex/papers/la_dir/l40n22u16.ps}}}}
\centerline{\hspace*{-3cm}
\mbox{
\rotate{
\epsfysize=8.2cm
\epsffile{/u/heinz/tex/papers/la_dir/l40n34u16.ps}}}}
\caption[f5]{\em ``$4k_F$'' singlet excitation spectrum for a Hubbard chain
of 40 sites with
22 and 34 electrons. The lowest ``arc'' from $k/\pi = 0.05$ to $k/\pi = 1.1$
(or $k/\pi = 1.7$) is obtained by varying $i_0$ at
fixed $I_0 = (N_0+1)/2$ (cf. eq. (\ref{si2})), the higher arches correspond
to increasing $I_0$ up to $(L-1)/2$.}
\label{fig5}
\end{figure}

\paragraph{``$\bf 4k_F$'' singlet states.}
Excited states are obtained by varying the quantum numbers. The first
possibility, giving rise to excited singlet states, is obtained by removing
one of the $I$'s from the ground state sequence (\ref{si1}) and adding a
``new'' $I$:
\begin{eqnarray}
\nonumber
\{ I_j \} & = & \{-(N_0-1)/2, \dots ,
-(N_0-1)/2+i_0-1, \\
\label{si2}
& & \quad \quad -(N_0-1)/2+i_0+1,\dots,(N_0-1)/2, I_0 \} \virg \\
\nonumber
\{ J_\alpha \} & = & \{-(N_0/2-1)/2, \dots , (N_0/2-1)/2 \} \virg
\end{eqnarray}
where $|I_0 | > (N_0-1)/2$. This is a two--parameter family of excited
states, called (somewhat misleadingly) ``particle--hole excitation'' by Coll.
To understand the excited
states, we shall in the following consider systems of finite size, rather
than
taking the thermodynamic limit directly. It should however be quite evident
how spectra like those of \fref{fig5} develop into a true continuum in
the thermodynamic limit. In
\fref{fig5} we show numerical results for the energy--momentum
spectrum of the states (\ref{si2})  for a chain of 40 sites.
The same states with $k \rightarrow -k$ are obtained using negative $I_0$.
One notices a sharp minimum in the excitation energy at $k/\pi = 1.1 (1.7) =
4k_F$ (this is why I call these excitations ``$4k_F$'' singlets).
In the thermodynamic limit, the gap at $4k_F$ vanishes. These excitations
are at the origin of the power--law behavior in the density--density
correlation function (\ref{nn}) around $4k_F$. Moreover, in the
bosonization formalism, the $4k_F$ density correlations are entirely
determined by the charge ($\phi_\rho$) modes. Consequently, we identify
the charge velocity $u_\rho$ as the slope of the excitation spectrum of
\fref{fig5} at $k=0$. Finally, we notice that the total number of states in
this branch decreases as one approaches half--filling, and that this branch
disappears altogether at half--filling. In fact, at half--filling, charge
excitations have a gap and are described by solutions with
complex $k$'s \cite{woynarovich_complexk}.

\begin{figure}
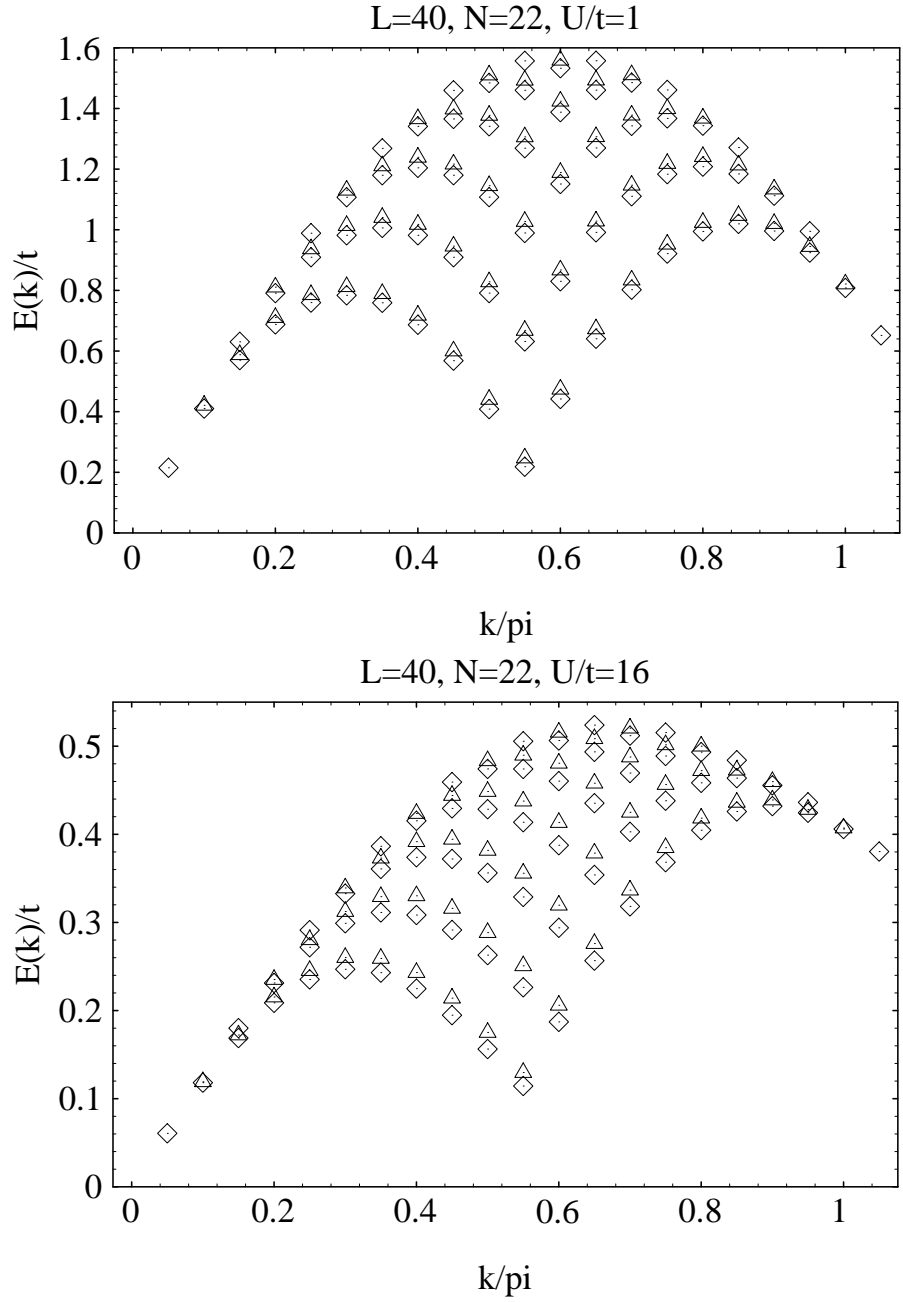

\centerline{\mbox{\rotate{\epsfysize=12cm
\epsffile{/u/heinz/hubbard-1d/hub0/st1.ps}}}}
\centerline{\mbox{\rotate{\epsfysize=12cm
\epsffile{/u/heinz/hubbard-1d/hub0/st16.ps}}}}
\caption[f6]{\em ``$2k_F$'' spin singlet ($\triangle$) and triplet
($\diamond$)
excitation spectrum for Hubbard chains of 40 sites with 22 electrons for
different interaction strengths.} \label{fig6}
\end{figure}

\paragraph{``$\bf 2k_F$'' triplet and singlet states.}
Excitations of the $J$'s with all $\lambda$'s and $k$'s real are only
possible if $M < N/2$. The simplest excitations of this type are obtained
considering $M = N/2 - 1$ which has total spin $S=1$ (triplet). Now the
restriction (\ref{restj}) allows $N/2 + 1$ different $J$'s, i.e. we have two
free parameters (the ``holes'' in the $J$--sequence), leading to sequences
of quantum numbers of the form
\begin{eqnarray}
\nonumber
\{ I_j \} & = & \{-N_0/2+1, \dots , N_0/2 \} \virg \\
\label{sj1}
 J_1  & = & -N_0/4 + \delta_{\alpha_1,1} \virg \\
\nonumber
 J_\alpha & = & J_{\alpha-1} + 1 + \delta_{\alpha,\alpha_1} +
\delta_{\alpha,\alpha_2} \quad (\alpha = 2,\dots,M) \virg
\end{eqnarray}
where $1 \le \alpha_1 < \alpha_2 \le
M$, and $\delta_{\alpha,\beta}$ is the usual Kronecker symbol.
Numerical results for this type
of excitations are shown in \fref{fig6}. Corresponding states with
negative $k$ are obtained shifting the $\{ I_j \}$ in (\ref{sj1}) by one
unit to the left. There now is a sharp minimum at
$k/\pi = 0.55 = 2k_F$, the gap again vanishing in the thermodynamic limit.
In the long--wavelength limit, these states are the only spin--carrying
excitations at constant particle number, so that the slope of the spectrum
at $k=0$ is equal to the spin velocity of the bosonized model. The
low--energy excitations around $2k_F$ are responsible for the spin
contribution to the $2k_F$ spin--spin correlations. As in
 \fref{fig5}, the structure of the excitations doesn't change much
between weak and strong correlations, however, the energy scale does. In
fact, the lowering of the energy scale in going to strong correlations
corresponds to the lowering of the exchange energy ($\approx 4t^2/U$) in the
strong correlation limit.
The results of \fref{fig6} show apparent gaps at $2k_F$ and $4k_F$.
These are finite size effects: in the thermodynamic limit $L \rightarrow
\infty$ the gaps vary like $1/L$, and simultaneously
the spectra develop into a continuum.

Together with the triplet excitations (\ref{sj1}) there are also singlet
states ($M = N_0/2$), which are obtained by having one pair of complex
conjugate $\lambda$'s among the solutions to the original equations
(\ref{be1}, \ref{be2}). The energies of these states are shown by triangles
in \fref{fig5}. It is remarkable that these states are nearly
degenerate with the triplet states and in fact become exactly degenerate in
the thermodynamic limit.

The existence of singlets and triplets with the
same energy shows that these states are in fact the combination of two {\em
noninteracting spin--1/2 objects}, commonly called {\em spinons}. Of
course,
because of total spin conservation, these objects can only be excited
in pairs as long as one keeps the total number of particles fixed.

\begin{figure}
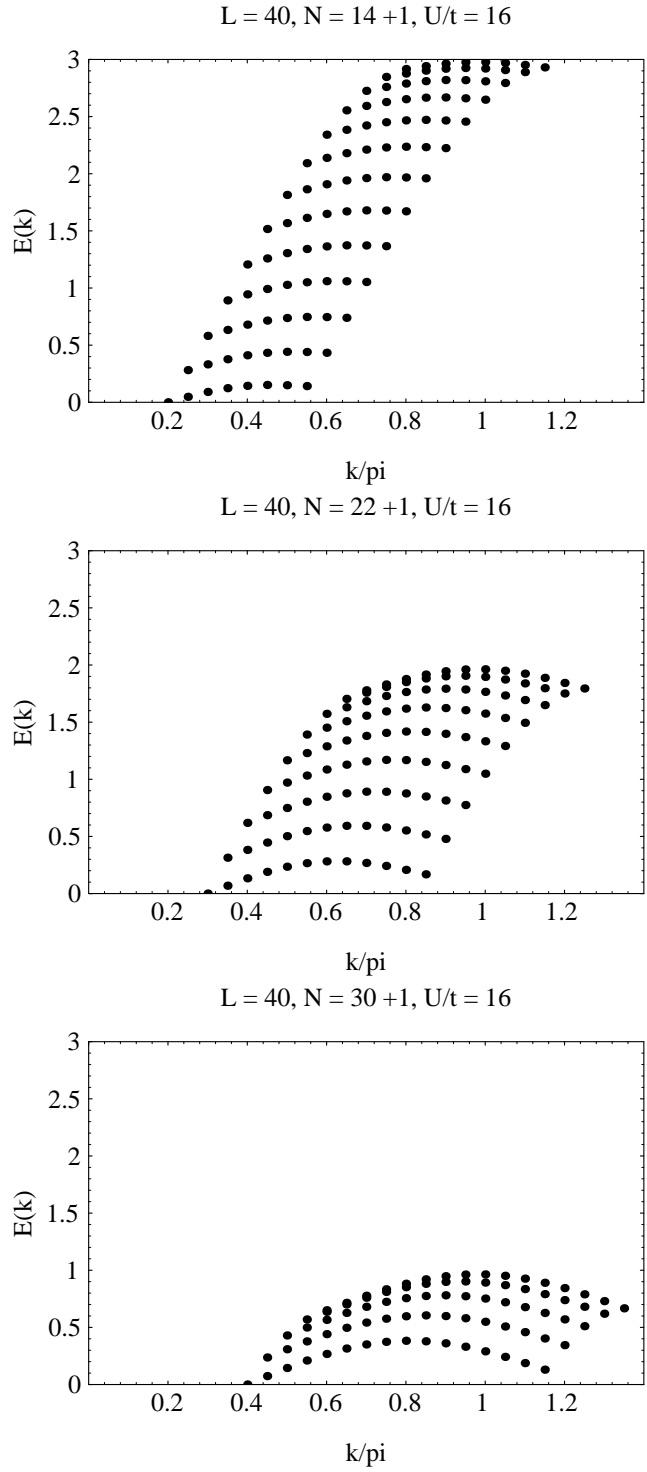

\centerline{\mbox{\rotate{\epsfysize=9cm
\epsffile{/u/heinz/hubbard-1d/hub3-res/l40n14u16.ps}}}}
\centerline{\mbox{\rotate{\epsfysize=9cm
\epsffile{/u/heinz/hubbard-1d/hub3-res/l40n22u16.ps}}}}
\centerline{\mbox{\rotate{\epsfysize=9cm
\epsffile{/u/heinz/hubbard-1d/hub3-res/l40n30u16.ps}}}}
\caption[f8]{\em
Excitation spectra for one particle added into a Hubbard chain of 40
sites with 14, 22, and 30 electrons. The shallow arches correspond to
varying $\alpha_1$ (cf. eq. (\ref{sj2})) at constant $I_0$, and $I_0$
increases from one arc to the next. Zero energy corresponds to the $N_0+1$
particle ground state.}
\label{fig8}
\end{figure}

\paragraph{Added particle.}
Adding one
particle to the $4\nu + 2$ ground state and leaving $M$ unchanged, the $I$'s
and the $J$'s are HOI. There are now $M+1$ allowed values for the $M$
distinct $J$'s, and the low-energy states then are parameterized by
\begin{eqnarray}
\nonumber
\{ I_j \} & = & \{-(N_0-1)/2, \dots , (N_0-1)/2 ,I_0 \} \virg \\
\label{sj2}
 J_1  & = & -M/2 + \delta_{\alpha_1,1} \virg \\
\nonumber
 J_\alpha & = & J_{\alpha-1} + 1 + \delta_{\alpha,\alpha_1}
 \quad (\alpha = 2,\dots,M) \virg
\end{eqnarray}
where $| I_0| > (N_0-1)/2$, and $1 \le \alpha_1 \le M$. Corresponding
spectra for different bandfillings are shown in \fref{fig8} for $I_0> 0$.
The symmetric spectra with negative $k$ are again obtained using $I_0 < 0$.
The state of minimal excitation energy has momentum $k_F$, as in the
noninteracting case. The shallow arches in \fref{fig8} correspond to
varying $\alpha_1$, and are in fact of the same shape as the lowest branch
(from $k = 0.05\pi$ to $k= 0.55\pi$) in \fref{fig6}, i.e. they
correspond to {\em single--spinon} excitations. Close to $k_F$ the energy
of theses states varies as $u_\sigma (k-k_F)$. On the other hand, varying
$I_0$, one goes from one arc to the next, and the corresponding
excitation energy (the upper limit of the quasi--continuum) varies as
$u_\rho (k-k_F)$. One also sees that
going from one arc to the next in \fref{fig8} (i.e. increasing $I_0$) the
shape of an individual arc is basically unchanged. Varying $I_0$ corresponds
to a variation of the momentum of the added particle, and the figure thus
shows that the total energy of a state is just the sum of the spinon energy
and the ``charge'' energy associated with the added particle. One thus sees
that {\em charge and spin degrees of freedom do not interact}. This is
certainly
in agreement with the predictions of the bosonization formalism, however,
the fact that spin and charge separate even in highly excited states
is special to the Hubbard model (a priori, the bosonized theory can only be
expected to be an effective low--energy theory for the Hubbard and other
lattice models).

\begin{figure}
\centerline{\mbox{\rotate{\epsfysize=10cm
\epsffile{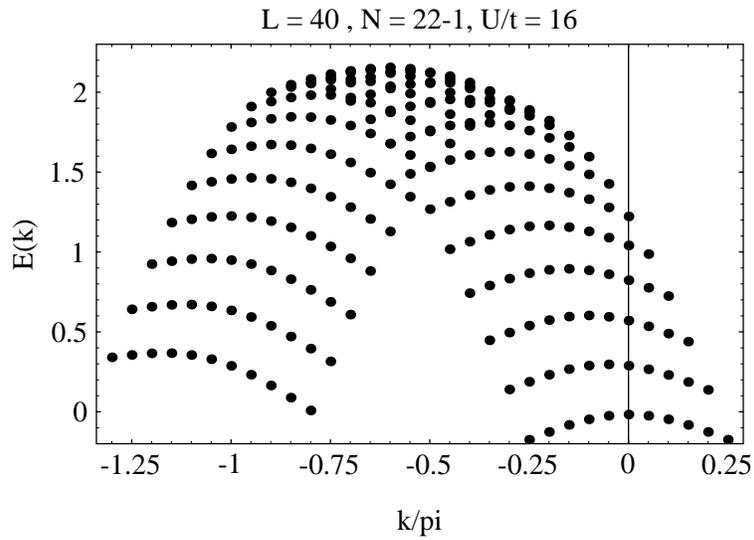}}}}
\caption[f8]{\em
Excitation spectra for an added hole in a Hubbard chain of 40
sites with 22 electrons. The shallow arches correspond to
varying $\alpha_1$ (cf. eq. (\ref{sj3}) at constant $j_1$, and $j_1$
increases from one arc to the next.}
\label{fig9}
\end{figure}

Another notable fact in \fref{fig8} is the number of available states as
$I_0$ is varied: for $N_0=14,22,$ and $30$ there are respectively $13,9,$
and $5$ spinon arches. This means that  without exciting the spins there
are $26,18,$
and $10$ states for the extra particle available (counting states at negative
$k$), i.e. the $I_0$ branch stops at $k=\pi - k_F$, rather than at $k=\pi$
as in a noninteracting system. The explanation of this fact is rather
straightforward for large $U$ when double occupancy of sites is forbidden:
in a system with $L$
sites and $N_0$ electrons, there are only $L-N_0$ sites available at low
energies. These states then form the ``band'' in  \fref{fig8}.
There are of course states involving doubly occupied sites, however these
are separated from the continuum of \fref{fig8} by a gap (these states
are solutions of (\ref{be1}, \ref{be2}) with complex $k$'s
\cite{woynarovich_complexk}).
This separation of states occurs for any, even very small $U$, and can
actually be shown by a perturbative argument \cite{ching_pert}.
\paragraph{Added hole.}
Finally, let us consider states with one hole in the $4 \nu + 2$
groundstate ($N= N_0-1, M = (N-1)/2$). Then both the $I$'s and the $J$'s are
integers.
The energy is minimized chosing consecutive $I$'s between $-(N-1)/2$ and
$(N-1)/2$, but there are $M+1$ possibilities for the $M$ $J$'s. States
corresponding to the sequence
\begin{eqnarray}
\nonumber
\{ I_j \} & = & \{-(N-1)/2, \dots , (N-1)/2  \} \virg \\
\nonumber
 J_1  & = & -M/2 + \delta_{\alpha_1,1} \virg \\
 J_\alpha & = & J_{\alpha-1} + 1 + \delta_{\alpha_1,\alpha}
 \quad (\alpha = 2,\dots,M)
\end{eqnarray}
are shown as the lowest arc between $k=-0.25 \pi$ and $k = 0.25 \pi$.
This is a one--spinon branch, this state having necessarily $S=1/2$, with
velocity $u_\sigma$. One can of course also create a hole in the sequence of
$I$'s. The energy spectrum for the quantum numbers
\begin{eqnarray}
\nonumber
 I_1  & = & -(N+1)/2 + \delta_{j_1,1} \\
 I_j  & = & I_{j-1} + 1 + \delta_{j_1,j} \quad
 (j = 2,\dots,N) \\
\label{sj3}
\nonumber
 J_1  & = & -M/2 + \delta_{\alpha_1,1} \virg \\
 J_\alpha & = & J_{\alpha-1} + 1 + \delta_{\alpha_1,\alpha}
 \quad (\alpha = 2,\dots,M)
\end{eqnarray}
(where the free parameters obey $1 \le j_1 \le N$, $1 \le \alpha_1 \le M$)
is also shown in \fref{fig9}. Similarly to \fref{fig8}, one notices that
varying the ``charge'' quantum number $j_1$ one creates a branch with
velocity $u_\rho$.

\subsection{Low energy properties of the Hubbard model}
\label{corrsec}
\subsubsection{Luttinger liquid parameters}
\label{lutt}
In a weakly interacting system the coefficients
$K_\rho$ and $u_\nu $ can be determined
perturbatively. For example, for the Hubbard model
one finds
\begin{equation}  \label{pert1}
K_\rho = 1 -  U/(\pi v_F) + ... \virg
\end{equation}
where $ v_F = 2t \sin(\pi n/2)$ is the Fermi velocity for $n$ particles
per site.
For larger $U$ higher operators appear in the continuum Hamiltonian
(\ref{hbos}), e.g. higher derivatives of the fields or cosines of
multiples of $\sqrt8 \phi_\sigma$.
These operators are irrelevant, i.e. they renormalized to zero and do not
qualitatively change  the
long-distance properties, but they do lead to nontrivial
corrections to the coefficients $u_\nu,K_\rho$.
In principle these corrections can be treated order by order in
perturbation theory. However, this approach is obviously unpractical for
large $U$, and moreover it is at least possible that perturbation theory is
not convergent. To obtain the physical properties for arbitrary $U$ a
different approach is therefore necessary.

I note
two points: (i) in the small-$U$ perturbative regime, interactions
renormalize to the weak-coupling fixed point $g_1^*=0, K_\sigma^* =1$;
(ii) the exact solution \cite{lieb_hubbard_exact} does not show any
singular behavior at nonzero $U$, i.e. large $U$ and small $U$ are the
same phase of the model, so that the long-range behavior even of the
large $U$ case is determined by the fixed point $g_1^* = 0$. Thus, the
low energy properties of the model are still determined by the three
parameters $u_{\rho,\sigma}$ and $K_\rho$.

The velocities $u_{\rho,\sigma}$ can be obtained from the long wavelength
limit of the ``$4k_F$'' and ``$2k_F$'' excitations discussed above. In the
thermodynamic limit the corresponding excitation energies are easily found
from the numerical solution of a linear integral equation
\cite{coll_excitations}.
Results are shown in \fref{ur} for various values of
$U/t$. Note that for $U=0$ one has $u_\rho = u_\sigma = 2t \sin (\pi
n/2)$, whereas for $U \rightarrow \infty$ $u_\rho = 2t \sin (\pi n)$,
$u_\sigma = (2 \pi t^2 /U) (1 -\sin (2\pi n)/(2\pi n))$. In the
noninteracting case $u_\sigma \propto n$ for small $n$, but for {\em
any} positive $U$  $u_\sigma \propto n^2$. The Wilson ratio,
eq.(\ref{rw}), obtained from the velocities is shown in \fref{ur}. For
$U=0$ one has $R_W = 1$, whereas for $U \rightarrow \infty$ $R_W = 2$ for $n
\neq 1$.

To obtain the parameter $K_\rho$ from the exact solution note that the
gradient of the phase field $\phi_\rho$ is proportional to the particle
density, and in particular a constant slope of $\phi_\rho$ represents a
change of total particle number. Consequently, the coefficient $u_\rho /
K_\rho $ in eq. (\ref{hnu}) is proportional to the variation of
the ground state energy $E_0$ with particle number
\cite{schulz_hubbard_exact}:
\begin{equation}
\label{kr}
\frac{1}{L} \frac{\partial^2 E_0(n)}{\partial n^2} = \frac{\pi}{2}
\frac{u_\rho}{K_\rho} = \frac{1}{n^2\kappa} \point
\end{equation}
Equation (\ref{kr}) now allows the direct determination of $K_\rho$:
$E_0(n)$ can be
obtained solving (numerically) Lieb and Wu's \cite{lieb_hubbard_exact}
integral equation, and
$u_\rho$ is already known. The
results for $K_\rho$ as a function of particle density are shown in
\fref{krho}{ } for different values of $U/t$.
For small $U$ one finds in all cases agreement with
the perturbative expression, eq. (\ref{pert1}), whereas for large $U$
$K_\rho \rightarrow 1/2$.
The limiting behavior for large $U$ can be understood noting that for
$U=\infty $ the charge dynamics of the system can be described by
noninteracting {\em spinless} fermions (the hard-core
constraint then is satisfied by the Pauli principle) with $k_F$ replaced
by $2k_F$. Consequently one finds a contribution proportional to $ \cos
(4k_F x) x^{-2}$ in the density-density correlation function, which from
eq. (\ref{nn}) implies $K_\rho = 1/2$. One then finds an asymptotic
decay like $\cos(2k_Fx) x^{-3/2} \ln^{1/2}(x)$ for the spin-spin
correlations, eq.(\ref{ss}), and an exponent $\delta=1/8$ in the
momentum
distribution function. The result $\delta = 1/8$ has also been found by
 Anderson and Ren \cite{anderson_ren_losal}, and by Parola and Sorella
\cite{parola_inf}. Ogata and Shiba's
numerical results \cite{ogata_inf} are quite close to these exact values.

As is apparent from \fref{krho}, the strong-coupling value $K_\rho = 1/2$
is also reached in the limits $ n \rightarrow 0,1$ {\em for any positive
$U$.} For $ n \rightarrow 0$ this behavior is easily understood: the
effective interaction parameter is $U/v_F$, but $v_F$ goes to zero in
the low-density limit (corresponding to the diverging density of
states). The limit $n \rightarrow 1$ is more subtle: in this case nearly
every site is singly occupied, with a very low density of holes. The
only important interaction then is the short range repulsion between
holes, which can be approximated by treating the {\em holes} as a gas of
spinless noninteracting fermions. Using (\ref{kr}), one then again finds
$K_\rho = 1/2$.

We note that in the whole parameter region, as long as the interaction is
repulsive one always has $K_\rho < 1$, which means that magnetic
fluctuations are enhanced over the noninteracting case. On the other hand,
superconducting pairing is always suppressed.

The results of \fref{krho} are valid
for $n \rightarrow 1$, but not for $n=1$. In the latter case,
there is a gap in the charge excitation spectrum, as expected from the
umklapp term (\ref{g3}), and the correlations of $\phi_\rho$ become long
ranged ($K_{\rho,eff}=0$). Close to half--filling, the asymptotic behavior
of correlation functions then exhibits a crossover behavior from
half--filled--like behavior at short distances to the general form at long
distances, as discusses in sec.\ref{halfsec} above (see in particular eq.
(\ref{cross}).

Results equivalent to the present ones can be obtained using the conformal
invariance of the Hubbard model \cite{frahm_confinv,kawakami_hubbard}.
These results have subsequently be generalized to the case with an applied
magnetic field \cite{frahm_confinv_field}.

\subsubsection{Transport properties}
The exact solution of Lieb and Wu can also be combined with the
long--wavelength effective Hamiltonian (\ref{hbos}) to obtain some
information on the frequency--dependent conductivity $\sigma (\omega)$.
On the one hand, from eq. (\ref{sig0}) there is a delta function peak at
zero frequency of weight $2 K_\rho u_\rho$. On the other hand, the
total oscillator
strength is proportional to the kinetic energy \cite{baeriswyl_optique}:
\begin{equation}
\sigma_{tot} = \int_{-\infty}^{\infty} \sigma (\omega) d\omega = -\pi
\langle H_{kin} \rangle /L \point
 \end{equation}
Thus, both the weight of the dc peak and the relative weight of the dc
peak in the
total conductivity can be obtained and are plotted in \fref{sig}. As
expected,
far from half--filling, all the weight in $\sigma_{tot}$ is in the dc
peak. For exactly half--filling the dc conductivity vanishes,
due to the existence of a gap $\Delta_c$ for charge excitations created
by
umklapp scattering, and all the weight is at $\omega > \Delta_c$. Fig.2
then shows that as $n \rightarrow 1$ umklapp scattering progressively
transfers weight from zero to high frequency. The crossover is very
sharp for small or large $U$, but rather smooth in intermediate
cases ($U/t \approx 16$). This nonmonotonic behavior as a function of
$U$ can be understood noting that initially with increasing $U$ umklapp
scattering plays an increasingly important role. Beyond $U/t \approx
16$, however, the spinless--fermion picture becomes more and more
appropriate, and at $U = \infty$ one again has all the weight in the dc
peak.
The linear vanishing of $\sigma_0$ as $n \rightarrow 1$ implies a
linear variation of the ratio $n / m^*$ with ``doping''. By the
thermopower--argument of sec.\ref{halfsec} carriers are
hole--like for $n<1$, and electron--like for $n>1$, provided one is close to
the transition. The thermopower of the one--dimensional Hubbard model as
well as the rather subtle crossover occuring in the vicinity of the
critical point $n=1$, $U=0$ have been analyzed in detail recently
\cite{stafford_scaling,stafford_thermop}.

\subsubsection{Spin--charge separation}
The Hubbard model also provides a rather straightforward interpretation
of the spin--charge separation discussed above. Consider a piece of a
Hubbard chain with a half--filled band. Then for strong $U$ there will
be no doubly--occupied sites, and because of the strong short--range
antiferromagnetic order the typical local configuration will be
$$
\begin{array}{c}
\cdots
\uparrow \downarrow \uparrow \downarrow \uparrow \downarrow
\uparrow \downarrow \uparrow \downarrow \uparrow \downarrow
\cdots
\end{array}
$$
Introducing a hole will lead to
$$
\begin{array}{c}
\cdots
\uparrow \downarrow \uparrow \downarrow \uparrow O
\uparrow \downarrow \uparrow \downarrow \uparrow \downarrow
\cdots
\end{array}
$$
and after moving the hole one has (note that the kinetic term in the
Hamiltonian does not flip spins)
$$
\begin{array}{c}
\cdots
\uparrow \downarrow O \uparrow \downarrow \uparrow
\uparrow \downarrow \uparrow \downarrow \uparrow \downarrow
\cdots
\end{array}
$$
Now the hole is surrounded by one up and one down spin,
whereas somewhere else there are two adjacent up spins. Finally, a few
exchange spin processes lead to
$$
\begin{array}{c}
\cdots
\uparrow \downarrow O \uparrow \downarrow \uparrow
\downarrow \uparrow \downarrow \uparrow \uparrow \downarrow
\cdots
\end{array}
$$
Note that the original configuration, a hole surrounded by {\em two} up
spins has split into a hole surrounded by antiferromagnetically
aligned spins (``holon'') and a domain--wall like configuration, two
adjacent up spins, which
contain an excess spin 1/2 with respect to the initial antiferromagnet
(``spinon'').
The exact solution by Lieb and Wu contains two types of
quantum numbers which
can be associated with the dynamics of the spinons and holons,
respectively.
We thus notice that spinons and holons
\cite{kivelson_holon,zou_holon} have a well-defined meaning in the
present one--dimensional case.

The above pictures suggest that, as far as charge motion is concerned,
the Hubbard model away from half--filling can be considered as a
one--dimensional harmonic solid, the motion of the holes providing for an
effective elastic coupling between adjacent electrons. This picture has been
shown to lead to the correct long--distance correlation functions for
spinless fermions
\cite{haldane_bosons,emery_jeru}. For the case with spin, this suggests
that one can consider the system as a harmonic solid with a spin at each
site of the elastic lattice (lattice site = electron in this picture).

Let us now show that this gives indeed the correct spin correlation
functions. In a continuum approximation, the spin density then becomes
\begin{equation}
 \ve{\sigma}(x) = \sum_m \ve{S}_m \delta(x-x_m) \virg
\end{equation}
where the sum is over all electrons. After a Fourier transformation of
the delta function the spin--spin correlation function becomes
\begin{equation}  \label{ss2}
\langle \ve{\sigma}(x) \cdot \ve{\sigma}(0) \rangle =
\frac{1}{(2\pi)^2} \int dq \, dq' \sum_{m,m'} e^{-iqx} \langle \ve{S}_m
\cdot \ve{S}_{m'} e^{i(qx_m + q'x_{m'})} \rangle \point
\end{equation}
 The exchange energy between adjacent spins is  always
antiferromagnetic, whether there is a hole between them or not, and
consequently the low--energy spin dynamics always is that of an
antiferromagnetic chain of localized spins.
Under the additional assumption that the spin--spin correlations on an
elastic lattice depend mainly on the average exchange constant and not so
much on the fluctuations induced by motion of the electrons, the average in
(\ref{ss2}) factorizes into separate spin and charge factors. Following the
hypothesis about harmonic motion of the electrons, we write $x_m = R_m +
u_m$, where $R_m = m/|1-n|$ is the average position of the $m$th electron
and $u_m$ the displacement with respect to this position. Note that
the ``harmonic solid hypothesis'' implies that $u_{m+1} -u_m$ is small, but
not necessarily $u_m$ and $u_{m+1}$ separately. In the averages over atomic
positions now all terms with $q \neq q'$ vanish, and one has
\begin{equation}             \label{ss3}
\langle \ve{\sigma}(x) \cdot \ve{\sigma}(0) \rangle \approx
\int dq \sum_{m,m'} e^{-iqx} \langle \ve{S}_m \cdot\ve{S}_{m'} \rangle
e^{iq(R_m - R_{m'})} \langle e^{iq(u_m-u_{m'})} \rangle \point
\end{equation}
The average over $u_m$ in (\ref{ss3}) has a power law behavior:
$$
\langle e^{iq(u_m-u_{m'})} \rangle \approx |m-m'|^{-\alpha(q)} \virg
$$
with $\alpha(q) \propto q^2$, i.e. it has a smooth $q$--dependence. On the
other hand, the long--distance behavior of the spin--spin correlations of
an antiferromagnetic spin chain
is \cite{luther_chaine_xxz,affleck_log_corr,singh_logs}
$$
  \langle \ve{S}_m \cdot\ve{S}_{m'} \rangle
\approx (-1)^{m-m'} |m-m'|^{-1} \ln^{1/2}|m-m'|  \point
$$
Therefore in (\ref{ss3}) the $q$--integration is dominated by terms with $q
\approx \pi (1-n) = 2k_F$. Replacing the weakly $q$--dependent
exponent $\alpha(q)$ by $\alpha(2k_F)$ one obtains
\begin{eqnarray}
\langle \ve{\sigma}(x) \cdot \ve{\sigma}(0) \rangle & \approx &
\int dq e^{-iqx} \sum_{m,m'} \langle \ve{S}_m \cdot\ve{S}_{m'} \rangle
e^{iq(R_m-R_{m'})}  |m-m'|^{\alpha(2k_F)} \\
& = & \cos (2k_Fx) x^{-1-\alpha(2k_F)} \ln^{1/2}(x) \point
\end{eqnarray}
With the identification $\alpha(2k_F) = K_\rho$, this is precisely the
result (\ref{ss}). We thus have shown that the spin--spin correlations of a
correlated electron system can in fact be understood as those of an
elastic
lattice of spins. In that picture, the motion of the holes then only
provides the effective elasticity for the lattice.

\subsubsection{The metal--insulator transition}
The one--dimensional Hubbard model is insulating for $n=1$, $U>0$, but
conducting in all other cases. Moreover, in agreement with the discussion of
section \ref{umsec} , the metal--insulator transition occurs in a
different
way according to whether the interaction strength is varied at constant
carrier density or whether one varies the carrier density at constant $U$.

It seems worthwhile here to compare the metal--insulator transition
occuring as $n \rightarrow 1$ with other scenarios for strongly
correlated fermion systems in higher dimension (see the review by Vollhardt
\cite{vollhardt_helium_revue}). In the ``nearly localized'' picture,
effective mass effects predominate and enhance both the specific heat and the
spin susceptibility. Consequently, the Wilson ratio ($1/(1+F_0^a)$ in Fermi
liquid language) remains nonzero as the metal--insulator is approached. On
the other hand, in the ``nearly ferromagnetic'' (or paramagnon) picture, only
the spin susceptibility is enhanced significantly, and therefore $R_W$ can be
much larger than unity. The behavior found here in the one--dimensional case
is quite different from both these scenarios: generally $R_W < 2$, and
approaching the metal--insulator transition $R_W \rightarrow 0$. This occurs
because generally an enhancement of the mass of the {\em charge} carriers
(i.e. a decrease of $u_\rho$) has no influence on the spin degrees of freedom
(see fig.1). This is rather straightforwardly understood in terms of {\em
spin--charge decoupling}, as explained in the previous section: charge and
spin excitations move nearly independently of each other, and in particular
the spin dynamics is determined by antiferromagnetic nearest--neighbor
exchange. In particular the spin susceptibility remains finite even when the
mass of the charge carrier approaches infinity.

Let us discuss the metal--insulator transition in more detail. The fact
that $u_\rho$ and $\sigma_0$ vanish linearly as $n \rightarrow 1$ seems
to be consistent with a divergent effective mass at constant carrier
density because $u_\rho \approx 1/m^*, \sigma_0 \approx n/m^*$. A
constant carrier density is also consistent with the fact that $k_F =
\pi n /2$ is independent of $U$. It is {\em not consistent} with the
hole--like sign of the thermopower as $n \rightarrow 1$ from below, nor
with the electron--like sign as $n \rightarrow 1$ from above: if the
carriers are holes, the carrier density is the density of holes: $n^* =
1-n$. Treating the holes as spinless fermions, as already mentioned
before, one expects $\sigma_0 \rightarrow 0$ because $n^*
\rightarrow 0$, and $\gamma \rightarrow \infty$ because the density of
states of a one--dimensional band diverges at the band edges.
This agrees with what was found explicitly in section \ref{lutt}.
What is not so easily understood in this picture is the fact that $k_F$
(i.e. the location of the singularity of $n_k$) is given by its
free--electron value $\pi n /2$, rather then being proportional to
$n^*$. One should however notice that $n_k$ is given by the
single--particle Green's function which contains both charge and spin
degrees of freedom. The location of $k_F$ then may possibly be explained
by phase
shifts due to holon--spinon interaction. This is in fact suggested by
the structure of the wavefunction of the exact solution
\cite{ogata_inf}.

The magnetic properties do not agree with what one expects from an effective
mass diverging as $n \rightarrow 1$: $u_\sigma$ and therefor $\chi$ remain
finite. Moreover, the NMR relaxation rate would have the behavior $1/T_1 =
\alpha T + \beta \sqrt{T}$, where the first (Korringa) term comes from
fluctuations with $q \approx 0$, whereas the second term comes from
antiferromagnetic fluctuations with $q \approx 2k_F$. None of these
properties is strongly influenced by the diverging effective mass observed
e.g. in the specific heat. This fact is of course a manifestation of the
separation between spin and charge degrees of freedom.

\subsubsection{Other models}
It is clearly interesting to go beyond the Hubbard model. A simple
generalization is  the ``extended Hubbard model'' which includes
a nearest--neighbor repulsion:
\begin{equation}  \label{uv}
H= -t\sum_{ i,s} (a^{\dagger}_{is} a_{i+1,s}
 +  a^{\dagger}_{i+1,s} a_{is} )
+ U \sum_i n_{i \up} n_{i \down}
+ V \sum_i n_i n_{i+1} \virg
\end{equation}
For this model, exact eigenvalues can not be obtained in the thermodynamic
limit.
The parameters in eq. (\ref{kr}) can however be calculated reliably for
finite systems \cite{mila_zotos}. In particular, at quarter filling
one finds a Luttinger--liquid ground state, with $K_\rho = 1/4$ at the
metal--insulator transition which occurs with increasing $V$, in agreement
with the discussion of sec. \ref{ocsec}.

 Exact exponents can be obtained for the
model (\ref{uv}) in the limit $U\rightarrow \infty$: then one has
effectively  spinless fermions (with $k_F \rightarrow 2k_F$) with
nearest neighbor interaction, a model which can be exactly solved
using the Jordan--Wigner transformation into the XXZ spin chain. In
particular, the $4k_F$--component of (\ref{nn}) is related to the
correlation function of $S_z$. From the known results
\cite{luther_chaine_xxz} one obtains, for a
quarter--filled band ($n  =1/2$), $K_\rho = 1/(2+(4/\pi )
\sin^{-1}(v))$, $u_\rho = \pi t \sqrt{1-v^2}/\cos^{-1}(v)$, with
$v=V/2|t|$.
Now $K_\rho < 1/2$ is possible. For $v > 1$ the system is in a
dimerized insulating state. Approaching the insulating state from
$v<1$ both $K_\rho$ and $u_\rho$ remain finite, i.e. $\sigma_0$ jumps
to zero at $v=1$. For $n \neq 1/2$ the parameters $u_\rho, K_\rho$ can
be obtained from numerical results \cite{haldane_xxzchain}. Quite
generally, one has
$K_\rho > 1/8$, but $K_\rho = 1/2$ for $n \rightarrow 0,1$, independent
of $v$.
On the other hand, $u_\rho \rightarrow 0$ as $n \rightarrow 1/2$
for $v > 1$, i.e. in that case the weight of the dc conductivity goes
to zero continuously, the point $(v,n) = (1,1/2)$ is thus highly
singular. The same type of singularity also occurs at $U=0, n=1$ in the
Hubbard model \cite{stafford_scaling}. Interestingly enough, one has $K_\rho
> 1$ if $V < - \sqrt2
|t|$, i.e. a finite amount of nearest--neighbor attraction is sufficient
to lead to divergent superconducting fluctuations even for infinite
on--site repulsion.
Also note that the singularities in $u_\rho$ and $K_\rho$
at $v=-1$ (attractive interaction) represent a point of phase
separation.

For the $t-J$ model, there is one exactly solvable point ($t=J$) where exact
exponents can be found using the Bethe ansatz \cite{kawakami_tj}. Away from
this point, eq. (\ref{kr}) has been used to obtain $K_\rho$ from numerical
data \cite{ogata_tj}. It is not easy to study the metal--insulator
transition occuring for $n \rightarrow 1$ numerically, however the
results are consistent with $K_\rho=1/2$ in this limit. For large $J$ there
is a phase with predominantly
superconducting fluctuations ($K_\rho > 1$).

\sectio{Conclusion}
In this paper, we have seen that using
the bosonization method and exact solutions,
one obtains a rather complete picture of many physical properties of
interacting one--dimensional fermions. Probably
the most important feature arising is the Luttinger liquid like behavior,
characterized by non--universal power laws, together with the separation of
the charge and spin dynamics. One should also notice that there are no
qualitative differences between  weak and strong correlation.
Metal--insulator transitions occur for
commensurate bandfillings, and in particular at half--filing, for repulsive
interactions. The transitions at varying particle density
show qualitatively different behavior according to whether the bandfilling
is an even or odd fraction.

Another type of metal--insulator transition, not discussed in detail here,
occurs in the presence of disorder (see ref. \cite{giamarchi_loc} and
references therein). In this case, as is well--known, in the absence
of interactions all states are localized. Repulsive interactions enhance
localization. On the other hand, sufficiently strong attraction can lead to
delocalization. In cases where there is no spin gap, the transition to the
delocalized conducting (in fact superconducting) state occurs at $K_\rho
>2$, and at the transition one has a {\em non--universal} value of the
correlation
exponent, satisfying only $K_\rho^* \ge 2$. In the case with spin gap, the
superconducting state only occurs for $K_\rho > 3$. In this case, at the
transition one has a {\em universal} value $K_\rho^* = 3$. One should notice
that these large values of $K_\rho$ in fact correspond to rather strong
attractive
electron--electron interactions, and it seems doubtful that this can be
realized in an experimental system.

Unambigous experimental observation of Luttinger liquid like behaviour and
of the associated metal--insulator transitions is made difficult by the fact
that all possible known candidates are in fact only {\em
quasi}--one--dimensional, e.g. they consist of a parallel arrangement of
conducting chains. At sufficiently low temperatures one thus always crosses
over into a regime of effectively three--dimensional behaviour,
characterized in particular by the occurence of different kinds of ordered
states. The specifically one--dimensional behaviour is thus not always
easily identified. Nevertheless, a number of very interesting experiments do
exist. An early and spectacular example is the observation of diffuse X--ray
scattering at wavevector $4k_F$ in the compound TTF--TCNQ \cite{pouget_4kf}.
In fact, in
a perturbative theory, no such scattering is expected. However, within  the
boson picture of the Luttinger liquid, such scattering is indeed expected
for strongly repulsive interaction \cite{emery_4kf,lee_etal_4kf} (see also
eq. (\ref{nn}) and the subsequent discussion). More recently, NMR data on
the series of compounds $\rm (TMTSF)_2 X$ have shown Luttinger liquid
like behavior \cite{wzietek_nmr}, and in particular power--law dependence
of the relaxation
rate on temperature. In some of these compounds umklapp scattering is
sufficiently strong to induce a (relatively small) Mott--Hubbard insulating
gap, and one can then observe a crossover between high--temperature metallic
and low--temperature insulating behavior at a temperature between 100K and
200K. The most recent observation concerns photoemission on $\rm (TMTSF)_2
PF_6$ \cite{dardel_photo}, where the spectral density does not show a Fermi
edge like behavior,
but is rather reminiscent of a power--law behaviors as expected in one
dimension (see eq. (\ref{spec})). However, the observed exponent is $\delta
\approx 1$, a value that would imply very strong electron--electron
repulsion. One then would expect strong effects of umklapp scattering, i.e.
typically a non--metallic conductivity. However, experiment shows good
metallic behavior in the temperature region concerned. The photoemission
results thus do not seem to be fully understood. It should be pointed out
here that the metal--insulator transitions observed always occur at constant
band--filling, upon varying temperature, pressure, or chemical composition
(which in the last two cases is equivalent to changing the interaction
strength). Unfortunately, it has up to now been impossible to dope organic
conductors in a sufficiently controlled way that would make the transition
occuring as a function of doping observable.

Anderson has
suggested that Luttinger liquid behavior might also occur in two
dimensions \cite{anderson_luttinger}, as well as in coupled chain systems
\cite{anderson_q1d}. Under which circumstances this suggestion is correct
does not currently clear. At least for the coupled--chain case,
Anderson's suggestion is in contradiction with standard scaling
\cite{schulz_trieste}
and renormalization group arguments \cite{bourbon_couplage,fabrizio_q1d}
which indicate that interchain hopping is a strongly relevant perturbation
and therefore most likelily will destroy the Luttinger liquid behavior.

\bigskip
\noindent
{\bf Acknowledgment} I'm grateful to colleagues in Orsay, in particular T.
Giamarchi, D. J\'erome and J.P. Pouget,
for many stimulating discussions on the subject of these
notes. This work was in part supported by CEE research contract no. CII/0568.


\newpage

\begin{figure}
\centerline{\rotate{\epsfxsize=9cm  \epsffile{/u/heinz/post/us.ps}}}
\caption[dummy]{\em The charge and spin velocities $u_\rho$ (full
line) and $u_\sigma$ (dashed line)
for the Hubbard model, as a function of the band filling for different
values of $U/t$: for $u_\sigma$ $U/t=1,2,4,8,16$ from top to bottom,
 for $u_\rho$ $U/t=16,8,4,2,1$ from top to bottom in the left part of
the figure. \label{ur}
}
\centerline{\rotate{\epsfxsize=9cm  \epsffile{/u/heinz/post/rw.ps}}}
\caption[dummy]{\em The Wilson ratio $R_W$
for the one--dimensional Hubbard model,
as a function of the band filling for different
values of $U/t$ ($U/t = 16 ,8, 4, 2, 1$ for the top to bottom curves).
\label{rwf} }
\end{figure}

\begin{figure}
\centerline{\rotate{\epsfxsize=9cm  \epsffile{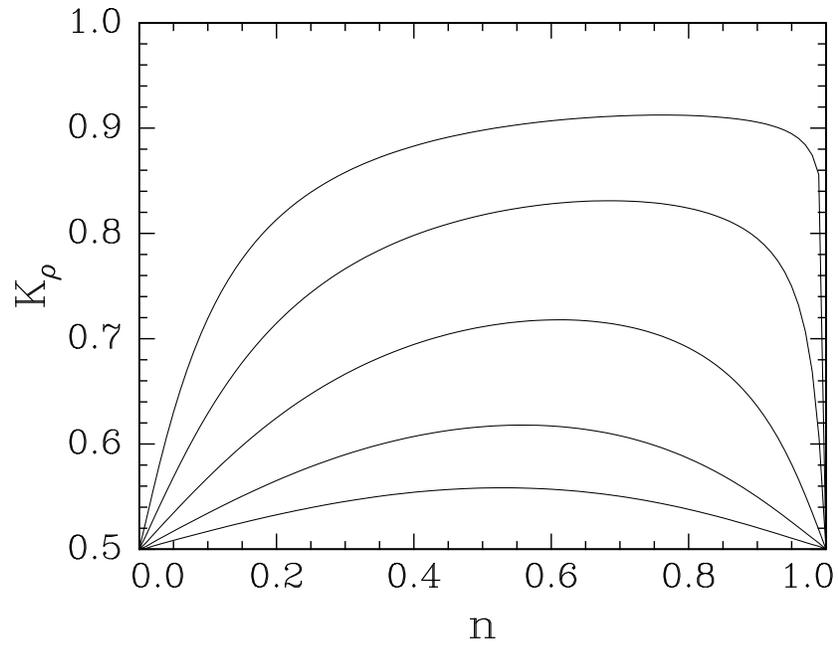}}}
\caption[dummy]{\em
The correlation exponent $K_\rho$
as a function of the bandfilling $n$ for different
values of $U$ ($U/t = 1,2,4,8,16$ for the top to bottom curves).
Note the rapid variation near $n=1$ for small $U$.}
\label{krho}
\end{figure}

\begin{figure}
\centerline{\rotate{\epsfxsize=9cm  \epsffile{/u/heinz/post/sig.ps}}}
\centerline{\rotate{\epsfxsize=9cm
\epsffile{/u/heinz/hubbard-1d/hubbet-res/sr.ps}}}
\caption[dummy]{\em {\rm Top:}
The weight of the dc peak in $\sigma (\omega )$ as a
function of bandfilling for different values of $U/t$
($U/t = 1,2,4,8,16$ for the top to bottom curves). \\
{\rm Bottom:}
Variation of the relative weight of the dc peak in the total
conductivity oscillator strength
as a function of the bandfilling $n$ for different values of $U$:
$U/t = 1$ (full line), $4$ (dashed), $16$ (dash--dotted), $64$
(dotted), and $256$ (dash--double-dotted).
}
\label{sig}
\end{figure}

\end{document}